\documentclass[sigconf]{acmart}

\usepackage{booktabs}
\usepackage{algorithm}
\usepackage{algpseudocode}
\usepackage{bbold}
\usepackage{enumitem}
\usepackage{subfigure}
\usepackage{colortbl}
\usepackage{setspace}
\usepackage{graphicx}

\newcommand\captionshrink{\vspace*{-0.5\baselineskip}}

\AtBeginDocument{%
  \providecommand\BibTeX{{%
    \normalfont B\kern-0.5em{\scshape i\kern-0.25em b}\kern-0.8em\TeX}}}

\copyrightyear{2022}
\acmYear{2022}
\setcopyright{acmcopyright}\acmConference[SIGIR '22]{Proceedings of the 45th International ACM SIGIR Conference on Research and Development in Information Retrieval}{July 11--15, 2022}{Madrid, Spain}
\acmBooktitle{Proceedings of the 45th International ACM SIGIR Conference on Research and Development in Information Retrieval (SIGIR '22), July 11--15, 2022, Madrid, Spain}
\acmPrice{15.00}
\acmDOI{10.1145/3477495.3531936}
\acmISBN{978-1-4503-8732-3/22/07}

\begin{document}

\title{Analyzing and Simulating User Utterance Reformulation in Conversational Recommender Systems}

\author{Shuo Zhang}
\authornote{The first two authors contributed equally.}
\affiliation{%
  \institution{Bloomberg}
  \city{London}
  \country{United Kingdom}
}
\email{imsure318@gmail.com}
\author{Mu-Chun Wang$^*$}
\affiliation{%
  \institution{University of Science and Technology \\ of China}
  \city{Hefei}
  \country{China}
}
\email{amoswang2000@mail.ustc.edu.cn}
\author{Krisztian Balog}
\affiliation{%
  \institution{University of Stavanger}
  \city{Stavanger}
  \country{Norway}
}
\email{krisztian.balog@uis.no}

\begin{abstract}
User simulation has been a cost-effective technique for evaluating conversational recommender systems.  However, building a human-like simulator is still an open challenge.  In this work, we focus on how users reformulate their utterances when a conversational agent fails to understand them.  First, we perform a user study, involving five conversational agents across different domains, to identify common reformulation types and their transition relationships.  A common pattern that emerges is that persistent users would first try to rephrase, then simplify, before giving up.  Next, to incorporate the observed reformulation behavior in a user simulator, we introduce the task of reformulation sequence generation: to generate a sequence of reformulated utterances with a given intent (rephrase or simplify).  We develop methods by extending transformer models guided by the reformulation type and perform further filtering based on estimated reading difficulty.  We demonstrate the effectiveness of our approach using both automatic and human evaluation.
\end{abstract}




\maketitle

\section{Introduction}
\label{sec:int}

A conversational recommender system (CRS) is a task-oriented dialogue system that supports its users in fulfilling recommendation-related intentions via a multi-turn conversational interaction~\citep{Jannach:2020:SCR}.  
A major differentiating factor from traditional recommender systems is that interactions are in natural language~\citep{Gao:2018:NAC}, which allows for a more natural elicitation of preferences~\citep{Radlinski:2019:CCP} and nuanced feedback on the suggested items~\citep{Balog:2021:OIM}.  Despite the tremendous advances made in NLP technology in recent years, natural language understanding continues to present a major challenge for CRSs~\citep{Jannach:2020:SCR,Gao:2018:NAC}.

\begin{figure}[t]
   \centering
   \includegraphics[width=0.45\textwidth]{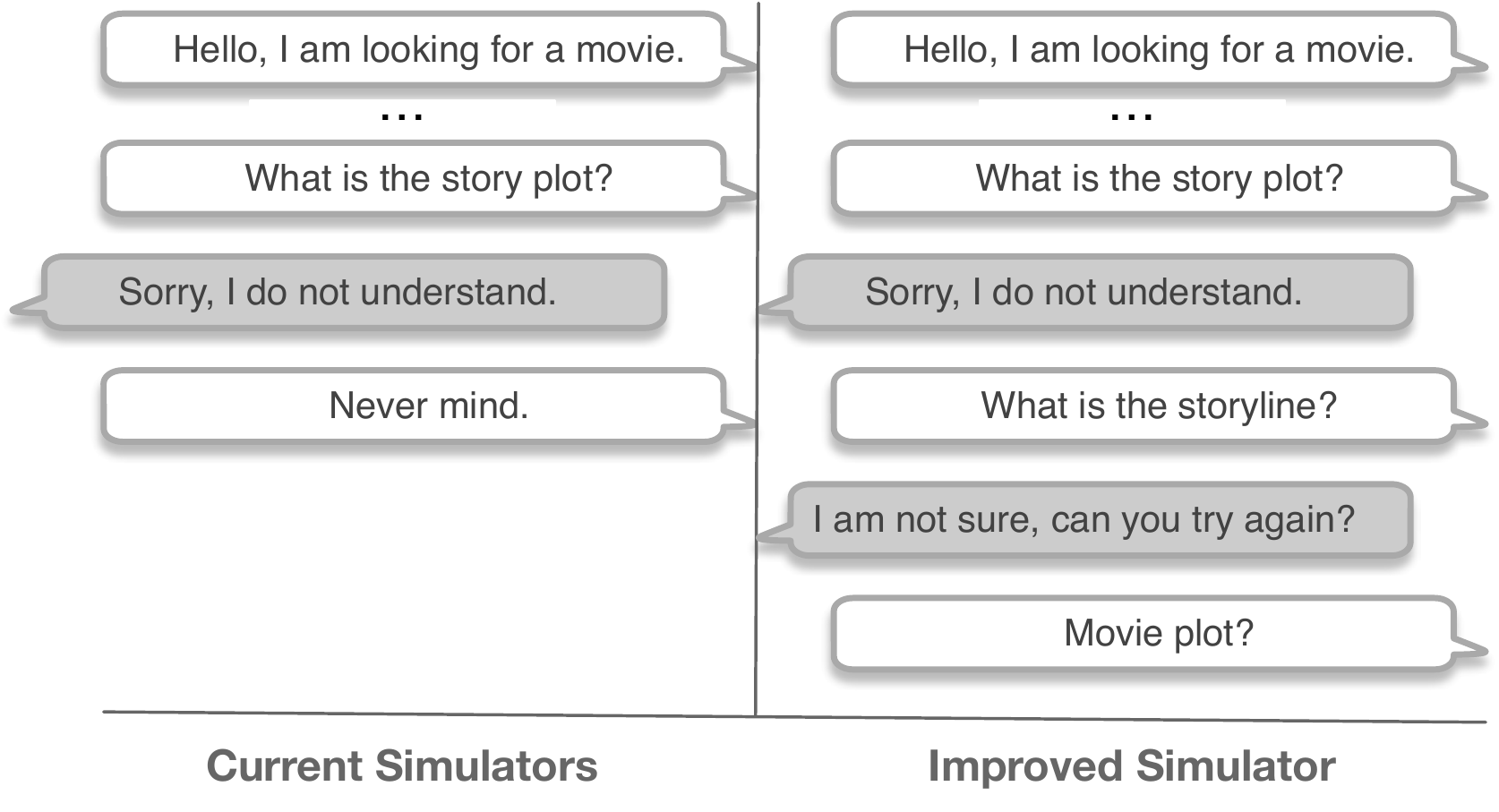} 
   \caption{Existing simulators (Left) will give up immediately when the agent fails to understand them. Our improved simulator (Right) makes multiple attempts by reformulating the request.}
  \vspace*{-1.5\baselineskip}
\label{fig:task}
\end{figure}
%

Progress in this area critically depends on the availability of appropriate evaluation methodology and resources.  Therefore, evaluation of conversational recommender systems, in and of itself, is a vitally important research topic~\citep{Jannach:2020:SCR,Gao:2018:NAC}. 
In general, recommender systems can be evaluated along various dimensions, including the quality of the recommendations, the perceived transparency of the process, or the ease-of-use of the system~\citep{Jannach:2020:SCR}.  While the same criteria can be also applied for conversational recommender systems, given their interactive nature, human-computer interaction aspects also need to be considered, related to the efficiency or quality of the conversation itself~\citep{Jannach:2020:SCR}.
End-to-end system evaluation is no longer possible using offline test collections, as due to its dynamic nature the conversation can branch ``infinitely'' at each user turn.  Therefore, offline test collections are limited to a single turn and a predefined set of possible user requests~\citep{Dalton:2020:TCC}.  Annotating entire dialogues using human judges is an alternative, but it is expensive, time-consuming, and does not scale~\citep{Zhang:2020:ECR}. 
A promising viable alternative is user simulation, which would provide a scalable and cost-efficient solution to automatic evaluation~\citep{Gao:2018:NAC,Balog:2021:SIGIRForum,Balog:2021:DESIRES}.

User simulation has been employed in the past in the context of task-oriented dialogue systems~\citep{Griol:2013:ADS,Schatzmann:2007:AUS,AsriHS:2016:ASM}.  There, simulation is primarily used for training reinforcement learning-based dialogue systems~\citep{Jan:2020:SEM, Huang:2020:KDB, Lei:2020:IPR}, and less so for evaluation.  Apart from a few noteable exceptions~\citep{Sun:2018:CRS}, those user simulators are relatively simplistic, using pattern-based replies. 
More recently, the idea of simulation has been revisited for conversational search and recommender systems, targeting more task-specific requirements.  Notably, \citet{Salle:2021:SEC} focus on generating answers to clarification questions, while \citet{Zhang:2020:ECR} develop models to guide recommendation-specific dialogue flows and to ensure the consistency of personal preferences.

A main shortcoming of existing user simulators is that they offer limited robustness when the generated user utterance is not understood by the system.  In the absence of a better option, they revert back to either restarting or terminating the dialogue.  Our objective is to develop a more human-like user simulator that would not ``give up'' immediately, if a given utterance was not understood by the system; see Figure~\ref{fig:task}.
The first main research question we ask is: \emph{How do users behave when their requests are not understood by a CRS?}
To answer this question, we design a user study and collect data using crowdsourcing on five conversational agents covering different domains, including movies, music, and travel, and annotate the dialogues using an established categorization of intents~\citep{Zhang:2020:ECR}.  We observe a typical behavioural pattern, according to which a persistent user would first try to rephrase, then to simplify their request, before giving up.

Motivated by these findings, we formulate a second main research question: \emph{How to model human-like reformulation behavior?}
To answer it, 
%
we introduce the task of \emph{reformulation sequence generation}: generating a sequence of different expressions for the same user intent, each with a specific reformulation type (rephrase or simplify).  
This task resembles the problems of question rewriting in conversational QA~\citep{Vakulenko:2020:QRF, Liu:2020:IUR, Su:2019:IMD, Aliannejadi:2019:ACQ} and query suggestions in conversational information-seeking tasks~\citep{Rosset:2020:LCS}.
However, reformulation sequence generation differs from question rewriting in major ways: as opposed to generating a single rewrite of the query, it outputs a diverse set of utterances corresponding to different reformulation types in the particular order in which they would be inputted by the user, should the system fail to understand them.

We propose to solve this task using a type-guided transformer, where the target reformulation type is provided as input.
While these models can generate different reformulations of the same utterance, not all are appropriate in a sense that they do not simplify the language.  Therefore, we additionally investigate a way to estimate reading difficulty and reject utterances that are deemed linguistically unacceptable.
This reformulation behavior can then be incorporated in the user simulation architecture via the response generation and natural language generation components. 

We evaluate our approach on the component-level, using both automatic and human evaluation.
Additionally, we perform contextual evaluation, by substituting human reformulation sequences with simulated ones in actual dialogues, and measuring to what extent human judges are able to distinguish between the two. 
We find that utterance reformulation contributes to making simulation more human-like, and a combination of type and reading difficulty can improve the utterance reformulation performance consistently across domains, methods, and evaluation measures.

In summary, the novel contributions of this paper are as follows.
\begin{itemize}[leftmargin=0.5cm]
    \item We conduct a user study to understand the traits of human utterance reformulation across five agents and three domains.
    \item We introduce the novel task of simulating utterance reformulation and explore how different transformer-based models can be used for addressing this task.
    \item We perform rigorous evaluation on the component-level as well as in an actual dialogue context.
    \item We construct a data collection consisting of dialogues by 150 human users, expert annotations for each dialogue turn, and 1.8k reformulated conversational utterances that may be used for training more realistic user simulators.
\end{itemize}
Resources are shared in the repository accompanying this paper: \url{https://github.com/iai-group/sigir2022-usersim}.

%
%
%

\section{Related Work}
\label{sec:rw}

This study lies in the intersection of conversational information access, evaluation, and utterance rewriting.

\subsection{Conversational Information Access} 
\label{sec:rw:cia}

\emph{Conversational information access} refers to a subset of task-oriented dialogue systems aimed at supporting users in fulfilling information-seeking goals via multi-turn conversations~\citep{Zhang:2020:ECR}. 
It includes the tasks of conversational search, QA, and conversational recommendations~\citep{Zamani:2022:FnTIR}.
Commercial products, like Amazon's Alexa, Google Assistant, Apple's Siri, Microsoft's XiaoIce, are assisting users in undertaking specific tasks in restricted domains, like restaurant reservations and music recommendation.
Recent studies have made progress in improving specific subtasks, including response generation, suggesting relevant queries~\citep{Rosset:2020:LCS}, asking clarifying question~\citep{Aliannejadi:2019:ACQ}, predicting user intent, and user preference modeling~\citep{Zhang:2020:ECR}.
\citet{Leif:2018:CAI} classify these supported actions and interactions identified in~\citep{Radlinski:2017:TFC} as \emph{query formulation}, \emph{set retrieval}, and \emph{mixed-initiative}. 
There also exists non-task-oriented dialogues systems, also known as chatbots, which support unstructured human-human interactions like chit-chat~\citep{Gao:2018:NAC}. 
However, we concentrate on the conversational recommender system in this work~\citep{Jannach:2020:SCR}.

\subsection{Evaluation}
Compared to traditional recommender systems, the evaluation of conversational recommender systems is focused more on the human-computer interaction aspects~\citep{Gao:2021:ACC}.
\citet{Jannach:2020:SCR} identify four main evaluation dimensions for conversational recommenders: (1) effectiveness of task support, (2) efficiency of task support, (3) quality of the conversation and usage, and (4) efficiency of the subtask. Each dimension considers different measurements on the component level and system level. 
For example, component measures for NLU include the accuracy, precision, and recall, while for NLG word-overlap based metrics are commonly employed~\citep{Belz:2006:CAA,Papangelis:2019:CMD}.  System-level metrics include reward and hit rate~\citep{Papangelis:2019:CMD, Peng:2018:DIP}.
Due to the dynamic nature of interactions, traditional (offline) evaluation methodology~\citep{Canameres:2020:OEO} is not suitable; alternative means of evaluation include user simulation and user studies.

\emph{User simulation} has been widely leveraged in the past for training the dialogue state tracking component of conversational agents using reinforcement learning algorithms, either via agenda-based or model-based simulation~\citep{Jannach:2020:SCR}.  The highly interactive nature of conversational information access systems has also sparked renewed interest in evaluation using user simulation within the IR community~\citep{Balog:2021:SIGIRForum,Salle:2021:SEC,Lipani:2021:TOIS,Sekulic:2022:EMC,Zhang:2020:ECR,Balog:2021:DESIRES}.
Recently, \citet{Zhang:2020:ECR} proposed a general framework for evaluating conversational recommender systems using user simulation. Their agenda-based simulator comprises NLU, NLG, and response generation components. 
However, building a human-like simulator is still an open challenge~\citep{Balog:2021:DESIRES}.
To bridge the gap, \citet{Salle:2021:SEC} focus on behavioral factors like cooperativeness and patience to build a more human-like simulator for information-seeking conversations.
\citet{Sekulic:2022:EMC} take this further by enabling simulated users to ask clarifying questions in a mixed-initiative setting.
\citet{Sun:2021:SUS} studied how to simulate user satisfaction in a human-like way for task-oriented tasks.
In this work, we concentrate on improving human-likeness in terms of how users would reformulate their utterances in case they are not understood by the system. 



The use of \emph{user studies} is a well-established human evaluation approach to understand user behavior in interactive tasks, e.g., search result selection~\citep{Vtyurina:2020:AMA}, query formulation~\citep{Trippas:2018:IDS}, and user engagement with virtual assistants~\citep{Trippas:2019:LAW, Vtyurina:2018:ERC, Vtyurina:2017:ECS,Qu:2018:ACU}. 
User studies may be categorized as lab, in-field, and crowdsourcing experiments~\citep{Jan:2020:SEM}.
For example,
\citet{Trippas:2018:IDS} conduct a lab-based user study to observe the characteristic of spoken interactions, such as how to initialize the query and reformulate them, and find that spoken conversational search is more complex and interactive than traditional search.
In~\citep{Trippas:2019:LAW} an in-field survey is performed with 401 respondents on daily tasks and activities with intelligent assistants in a work setting, which helps to understand the role of these assistants in supporting people engaged in work tasks.
The MSDialog dataset~\citep{Qu:2018:ACU} contains conversation data collected using crowdsourcing.
\citet{Qu:2018:ACU} find that short conversation cues such as ``Okay'' and ``Yup'' take place frequently in guided tasks with virtual assistants. 
\citet{Vtyurina:2018:ERC} carry out a Wizard-of-Oz study to study the role of implicit cues and conclude that a conversational system that is able to recognize and act upon such cues would enable its users to converse using a more natural style of language.


\subsection{Utterance Reformulation}
\emph{Utterance reformulation} or \emph{utterance rewriting} is related to the well-studied problem of query reformulation in IR~\citep{Rha:2017:ERQ}, and has been explored in many conversational QA tasks, e.g., for ambiguous questions~\citep{Vakulenko:2020:QRF},
incomplete utterances~\citep{Liu:2020:IUR}, multi-turn dialogue modeling~\citep{Su:2019:IMD}, and query clarification~\citep{Aliannejadi:2019:ACQ}.
Specifically, \citet{Vakulenko:2020:QRF} propose to reformulate ambiguous questions, whose reference information is only introduced earlier in the conversation, into implicit ones based on the context for the task of conversational QA. 
\citet{Liu:2020:IUR} consider incomplete utterance rewriting as a semantic segmentation task, casting it as a word-level edit matrix construction problem.
To improve the performance of multi-turn dialogue modeling, \citet{Su:2019:IMD} simplify the multi-turn into a single-turn task as a pre-process through utterance rewriting, which includes recovering all co-referred and omitted information. Based on a purpose-built 200k multi-turn conversational dataset, a pointer-generator model is trained as a rewriter for this task, which can be integrated into online chatbots.
\citet{Aliannejadi:2019:ACQ} propose to ask clarifying questions when users fail to formulate their complex information needs in a
single query. Similarly, \citet{Rosset:2020:LCS} suggest useful questions in conversational search to improve the users' engagement.
To enhance the diversity and quality of system responses, \citet{Lippe:2021:DTD} the combine two commonly used response generation methods, template-based and corpus-based, by paraphrasing template-based responses. 
In this paper, we focus on a particular reason why reformulation happens: when the conversational agent fails to understand the user utterance. We aim to model this reformulation behavior as a sub-component of the user simulator architecture.

\section{Analyzing Utterance Reformulations}
\label{sec:aur}



Most conversational recommender systems are still in an early stage of development, and many questions have not been well studied, including user behavior on utterance formulation~\citep{Trippas:2018:IDS}.
We carry out a user study 
to answer the following question: \emph{How do users behave when their requests are not understood by a conversational recommender system}?
To ensure that our observations are based on realistic user behavior, we need users to interact with an actual agent, as opposed to Wizard-of-Oz style studies, with an idealized agent behavior.
We design a crowdsourcing experiment on Amazon MTurk\footnote{\url{https://www.mturk.com/}} that involves 150 participants engaging with five conversational agents.
Based on the collected dialogues, we find that utterance rephrasing and simplification are the two most frequent types of behavior, and they take place equally frequently when interacting with a conversational recommender.


\subsection{Experimental Design and Participants}

\subsubsection{Objectives and Design Decisions}
Our main goal with this experiment is to understand how humans reformulate utterances when a conversational agent fails to understand them.  To ensure that our findings are not specific to a particular agent, but are generalizable observations, we consider a number of different agents.  

Another design decision we make is instructing the study subjects to keep communicating with the conversational system until they get a satisfactory recommendation. This is to mimic the behavior of a persistent user, which clearly introduces some bias. However, we consider this behavior as corresponding to maximum persistence, whereas the other extreme would be giving up immediately. The degree of user persistence can be adjusted and incorporated as a parameter in the simulator (see, e.g.,~\citep{Salle:2021:SEC}).



\subsubsection{Conversational Agents}
We carefully choose five existing third-party conversational agents that each of the systems should at least support the query formulation and set retrieval functionalities (cf. Sect.\ref{sec:rw:cia}). Three of these are for movie recommendation, one for music, and one for the travel domain. 
%
\begin{itemize}[leftmargin=0.5cm]
    \item \emph{\textbf{A1}: And chill}~\footnote{\url{http://www.andchill.io/}} is a movie theme chatbot that interacts on Facebook and provides Netflix recommendations. After answering a few questions such as a liked movie and the reason why liking it, the agent sends movie recommendations based on users' preferences.
    \item \emph{\textbf{A2}: Jarvis}~\citep{Zhang:2020:ECR} is a movie recommender system that is based on the Plato Research Dialogue System.\!\footnote{\url{https://github.com/uber-archive/plato-research-dialogue-system}} This agent can answer questions about movies such as directions, and ratings, and provide recommendations based on genres. It also solicits feedback on movies the user has watched.
    \item \emph{\textbf{A3}: IAI Movie Bot}~\citep{Habib:2020:IMA} is an open-source movie recommender system with a multi-modal chat interface. It focuses on modeling users' preferences dynamically via multi-turn dialogue.
    \item \emph{\textbf{A4}: Eddy Travel Bot} is a travel assistant that can help users organizing their trip on Telegram.\footnote{\url{https://t.me/eddytravels_bot}} Based on the travel destination, the agent gives recommendations, including flights, hotels, and restadeurants.
    \item \emph{\textbf{A5}: VKM Bot} is a music bot on Telegram that helps users find, listen, and download songs.\footnote{\url{https://telegram.me/vkm_bot}} The agent recommends a list of songs based on a song or artist name provided by the user. 
\end{itemize}
\emph{A2} and \emph{A3} are two similar agents in terms of supported functionalities. However, they are built on different dialogue frameworks. Besides, \emph{A2} is a conversational agent without ``keyboard guidance'' (i.e., buttons with suggestions) for users while \emph{A3} takes more initiative and guides users by offering them a set of options on buttons. We observe that 
users reformulate less with keyboard-based agents. 

\begin{table*}[t]
	\caption{Types of reformulation behavior in conversational recommender systems, where the agent fail to understand/satisfy the user. Each example lists three consecutive utterances, where two are from the user (U) and one by the agent (A).}
	\captionshrink
	\small
	\begin{tabular}{lll}
	\toprule
	\textbf{Type} & \textbf{Definition} & \textbf{Example} \\
	\midrule
	\emph{\textbf{Start/restart}}
		& Start of articulating an information need
		&  \textbf{U}: \emph{..., but I'm looking for a movie I've never seen.}  \\
		& (i.e., set of requirements)
		& \textbf{A}: \emph{That's all I got at the moment.} \\
		&
		& \textbf{U}: \emph{Dumb and Dumber is my favorite comedy...} \\
	\midrule
	\emph{\textbf{Repeat}}
		& Repeat the last utterance without 
		& \textbf{U}: \emph{I am looking for amazon premium movie.} \\ 
		& significant change
		& \textbf{A}: \emph{I'm pretty solid on a bunch of things so far.} \\
		&
		& \textbf{U}: \emph{I am looking for amazon premium movie.} \\
	\midrule
	\emph{\textbf{Repeat/rephrase}}
		& Repeat the last utterance using different
		& \textbf{U}: \emph{No, I'm looking for a restaurant.} \\ 
		& words/expression (about the same complexity)
		& \textbf{A}: \emph{Sorry, I didn't get that. Can you rephrase?} \\
		&
		& \textbf{U}: \emph{What restaurants do you know of in Fort Lauderdale?} \\
	\midrule
	\emph{\textbf{Repeat/simplify}}
		& Repeat the last utterance a simpler/shorter 
		& \textbf{U}: \emph{Never mind. Can you book me a taxi from the airport?} \\ 
		& words/expression (reduced complexity)
		& \textbf{A}: \emph{Sorry, I didn't get that. Can you rephrase?} \\
		&
		& \textbf{U}: A cab. \\
	\midrule
	\emph{\textbf{Clarify/refine}}
		& Clarify or refine the constraints 
		& \textbf{U}: \emph{Something light-hearted.} \\ 
		& expressed in the last user utterance
		& \textbf{A}: \emph{Can you tell me about a different one you like?} \\
		& 
		& \textbf{U}: \emph{I want a funny comedy that is light hearted.} \\
	\midrule
	\emph{\textbf{Change}}
		& Change the constraints (adding or removing) 
		& \textbf{U}: \emph{I don't think either of those are romantic comedies.} \\ 
		& expressed in the last user utterance
		& \textbf{A}: \emph{I think you should give Balls Of Fury a shot!} \\
		& 
		& \textbf{U}: \emph{Okay, let's try a movie that is similar to Speed.} \\
	\midrule
	\emph{\textbf{Stop}}
		&  End the session
		&  \textbf{U}: \emph{I give it a year looks really good, why did you recommend that one?} \\
		& 
		& \textbf{A}: \emph{Great! I'm really excited. These movies are going to be phenomenal!}
		\\
		&
		& \textbf{U}: \emph{Thanks again, I think I watch I Give it a Year.} \\
		
	\bottomrule
	\end{tabular}
    \label{tbl:reformulate}
\end{table*}
%

\subsubsection{Crowdsourcing Experiment}
The user study comprises of two parts: demographic information collection and dialogue collection. 
To learn how user behavior changes across user groups, we obtain demographic information by designing a small questionnaire, where we ask about the participants' background, such as age, gender, and education, and whether they have experience with conversational agents. 
%
%
We invite crowd workers on Amazon MTurk as participants and provide incentive payment for each task (\$2.5 for each task). 
Each participant is instructed to interact with a conversational recommender system either on Facebook or Telegram. 
They need to engage with the agent to find a liked movie, song, or travel plan.
They were given the following instructions:\footnote{The instructions were piloted on a smaller-scale lab-based study first (involving five undergraduate computer science students) to ensure that they are understandable.}\\

\hangindent=0.2cm
\hangafter=0
\noindent
    \emph{In this task, you are going to interact with a chatbot that will help you find movies that you might like. Your task is to engage in a conversation with this bot until you receive a recommendation for an item that you haven’t seen before but you’d be interested in watching. Note that the bot may not understand everything you say, in which case you’ll need to try again by stating your request somewhat differently.}\\
%

\noindent
For each of the agents, we invite around 60 participants to complete the task. 
We manually choose 30 questionnaires per agent after quality control\footnote{We exclude dialogues with very few turns, failure in asking for recommendations and repetitive records.} and select a total of $5 \times 30 =150$ dialogues (that is, by 150 unique users), which will be analyzed.
Of the 150 participants, 42\% are under 30 years old, 43.3\% are between 30 and 40, and 14.7\% are older than 40.
In terms of gender, 46\% are female and 54\% are male.
38.7\% of the annotators do not have higher education, 44.7\% have bachelor or college study, 14.7\% have a master degree, and 2\% have doctoral education.
Most of the annotators have prior experiences with using conversational chatbots such as Google Assistance, Siri, and Amazon Alexa, for finding restaurants, music, shopping, sports, or movies. However, 37\% have never asked for recommendations.

\subsection{Understanding Logged Dialogues}

To distill the utterance reformulation behavior from the logged dialogues, expert annotations were performed (by two of the paper's authors) on the exchange level along three dimensions: intent, slot, and reformulation.\footnote{Disagreements during the annotation process were resolved via discussion.}
In addition to facilitating user behavior analysis, these annotations also enable the use of the collected dialogues as training data for user simulation in the future.

\subsubsection{Intent Annotation}
\label{sec:uld:ia}
Intent reflects the action or interaction of the current utterance. 
We take the actions and interactions identified in~\citep{Zhang:2020:ECR} as intents, which are classified more broadly in~\citep{Radlinski:2017:TFC} as \emph{query formulation}, \emph{set retrieval}, and \emph{mixed-initiative}.
In the phase of query formulation, the system will \emph{elicit} the user to \emph{reveal} information need, which is further classified as \emph{disclose}, \emph{non-disclose}, \emph{revise}, \emph{refine}, and \emph{expand}. \emph{Set retrieval} is to display search and recommendation results, and it includes \emph{inquire} (\emph{list}, \emph{compare}, \emph{subset}, \emph{similar}) and \emph{navigate} (\emph{repeat}, \emph{back}, \emph{more}, \emph{note}, \emph{complete}).
\emph{Mixed initiative} include \emph{interrupt} and \emph{interrogate}.

\subsubsection{Slot Annotation}
In conversational recommendations, a slot refers to the variable for categorizing and interpreting the utterances in both the preferences and recommendations~\citep{Gao:2018:NAC}. For example, a movie mention or genre in \emph{disclose} reflects the user's tastes and preferences. 
We annotate specific items (movie, location, restaurant, hotel, song, musician) and movie genres.


\begin{table}[t]
\small
    \centering
        \caption{Agent intents that led to reformulations with relative frequencies (Ratio) and standard dev. across agents ($\sigma$).}
        \captionshrink
    \begin{tabular}{lcc|lcc}
    \toprule
    \textbf{Agent Intent} & \textbf{Ratio} & $\sigma$ & \textbf{Agent Intent} & \textbf{Ratio} &  $\sigma$ \\
    \midrule
    Failed & 0.62 & 0.0021 & Similar & 0.02 & 0.0013\\
    Suggest & 0.19 & 0.0015 & Repeat & 0.02 & 0.0021\\
    Elicit & 0.06 & 0.0017 & Non-disclose & 0.02 & 0.0018\\
    Extract & 0.03 & 0.0017 & End-disclose & 0.01& 0.0015\\
    List & 0.03 & 0.0015 & Clarify & 0 & 0\\
    \bottomrule
    \end{tabular}
    \label{tbl:RQ1}
\end{table}

\subsubsection{Reformulation Annotation}
\label{sec:aur:ra}

We assume that reformulation is likely to happen when the agent fails to understand or misunderstands the user. 
Based on the logged dialogues, we develop a new coding scheme, with a particular focus on capturing reformulation behavior; see the possible types and examples in Table~\ref{tbl:reformulate}. 
We annotate our dialogue collection with these reformulation types.

According to the reformulation types listed in Table~\ref{tbl:reformulate},
\emph{Start/restart} indicates the start of articulating an information need (i.e., set of requirements). In the example, the agent fails to recommend more items, thus the user restarts the dialogue by disclosing extra preferences.
When the agent fails to understand the user or simply gives an irrelevant reply, the user might \emph{repeat} the last utterance without significant change.
\emph{Repeat} can take place by repeating the last utterance using different words/expression (about the same complexity), namely \emph{Repeat/rephrase}, or repeating the last utterance a simpler/shorter words/expression with reduced complexity (\emph{Repeat/simplify}). 
When the agent does not completely understand the user, the user may clarify or refine the constraints expressed in their last utterance, denoted as \emph{Clarify/refine}. Under similar circumstances, the user might change the constraints (adding or removing) expressed in the last user utterance, which is defined as \emph{Change}. Finally, the user might \emph{End} the session once the information needs have been met, though the agent response is not as expected.

\subsection{Analyzing Reformulations}
Next, we analyze the annotated dialogues with the aim to gain an understanding of humans' reformulation behavior.\\

\noindent
\emph{(\textbf{RQ1}): When do reformulations occur?}
We focus on the last agent utterances that led to users' reformulations, making use of the intent annotations in Sect.~\ref{sec:uld:ia}.  Table~\ref{tbl:RQ1} displays the ratio of these intents. 
We find that reformulation takes place mostly when encountering \emph{Failed} replies, i.e., when the agent fails to understand the user. 
But it also appears with other system intents, such as \emph{Suggest} and \emph{Elicit}.
There are some indications of differences when considering demographic information, such as gender and age. 
For example, when age is over 40, the ratio of \emph{Failed} intent has dropped to 45\%, while \emph{Elicit} has raised to 23\%. 
The reason behind this might be that users' desire to respond changes with age. 
Younger users are more proactive in interacting with the agents when the agent failed. 
Older users, on the other hand, seem to be more patient with other intents, e.g., witnessed by a high ratio for reformulation in \emph{Elicit}.
We additionally investigate the ratios for each conversational agent and find that their distributions are similar to each other, witnessed by the low standard deviations.   
Besides, the distributions of agent intent are similar across different ages, gender, and experiences.
We note that these observations are indicative but not conclusive, as our data sample is limited in size. A larger-scale analysis of demographic differences is left for future work. \\


\noindent
\emph{(\textbf{RQ2}): Do people who have experience with conversational agents show different reformulation behavior from those that do not have that?} 
For each intent, we calculate the distributions of reformulation types for users both with and without experience with conversational recommendations.
For each pair of distributions, we first take the Levene test~\citep{Morton:1974:RTE} to examine the homogeneity of variance between them for each reformulation type. After this, we perform a t-test based on the mean values and find that the all the p-values are much larger than 0.1, meaning these two groups do not have differently significant behavior with regards to these intents.
These statistics are based on aggregated data across all agents, but we find that the same observations also hold for individual agents. \\

\begin{table}[!h]
    \centering
        \caption{Top reformulation patterns in our dataset.}
        \captionshrink
    \begin{tabular}{lcc}
    \toprule
    \textbf{Intent} & \textbf{Ratio} & \textbf{Count}\\ 
    \midrule
    Rephrase-Simplify & 0.13 & 80 \\ 
    Rephrase-Refine & 0.10 & 63 \\  
    Simplify-Rephrase & 0.08 & 53\\ 
    Refine-Simplify & 0.03 & 20\\ 
    Simplify-Refine & 0.03 & 19\\ 
    Refine-Rephrase & 0.03 & 18\\  
    \bottomrule
    \end{tabular}
    \label{tbl:reform}
\end{table}
%

\begin{figure*}[t]
   \centering
   \includegraphics[width=0.8\textwidth]{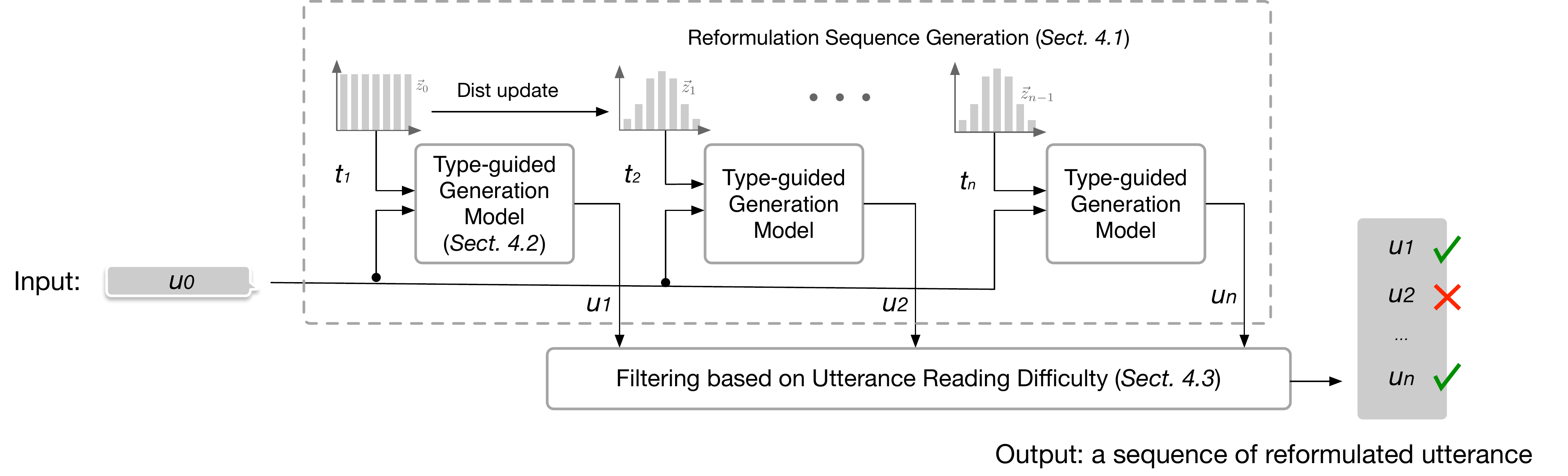}
   \caption{Reformulation sequence generation architecture.}
\label{fig:seq+smooth}
\end{figure*}

\noindent
\emph{(\textbf{RQ3}): Can reformulation patterns be identified?}
%
%
The sequential reformulations reveal how reformulation behavior changes when encountering continuous dissatisfying system replies.
To answer RQ3, we calculate the transition probabilities of the reformulation states. 
We split the logged dialogues into dialogue pieces by intents, where each reformulation is a state. We tracked the utterance by intent and slot; if the slot and intent have changed, we consider it as another dialogue piece.
One dialogue piece might occur in multiple states as the state changes upon the change of reformulation type.
For example, the user might reformulate differently for the same intent (in the same dialogue piece).
We construct the transition matrix $M$ based on the maximum likelihood estimation method~\citep{Iuliana:2009:MLE}. 
In particular, we consider the state transition sequence of each agent as a Markov process, and further maximize the likelihood for each element:
\begin{equation*}
m_{ij} = p(t_r=j|t_{r-1}=i) = \dfrac{\sum_{r} \mathbb{1}(t_{r-1}=i \land t_r=j)}{\sum_{r} \mathbb{1}(t_{r-1}=i)},
\label{RQ3_2}
\end{equation*}
where $t$ refers to the reformulation type, and $r$ means the current conversation turn. 
The frequencies of top transition patterns are reported as in Table~\ref{tbl:reform}.
For consecutive reformulations, we observe a typical pattern where a persistent user would first try to \emph{rephrase}, then \emph{simplify} before giving up. 

\begin{figure}[t]
   \centering
   \includegraphics[width=0.5\textwidth]{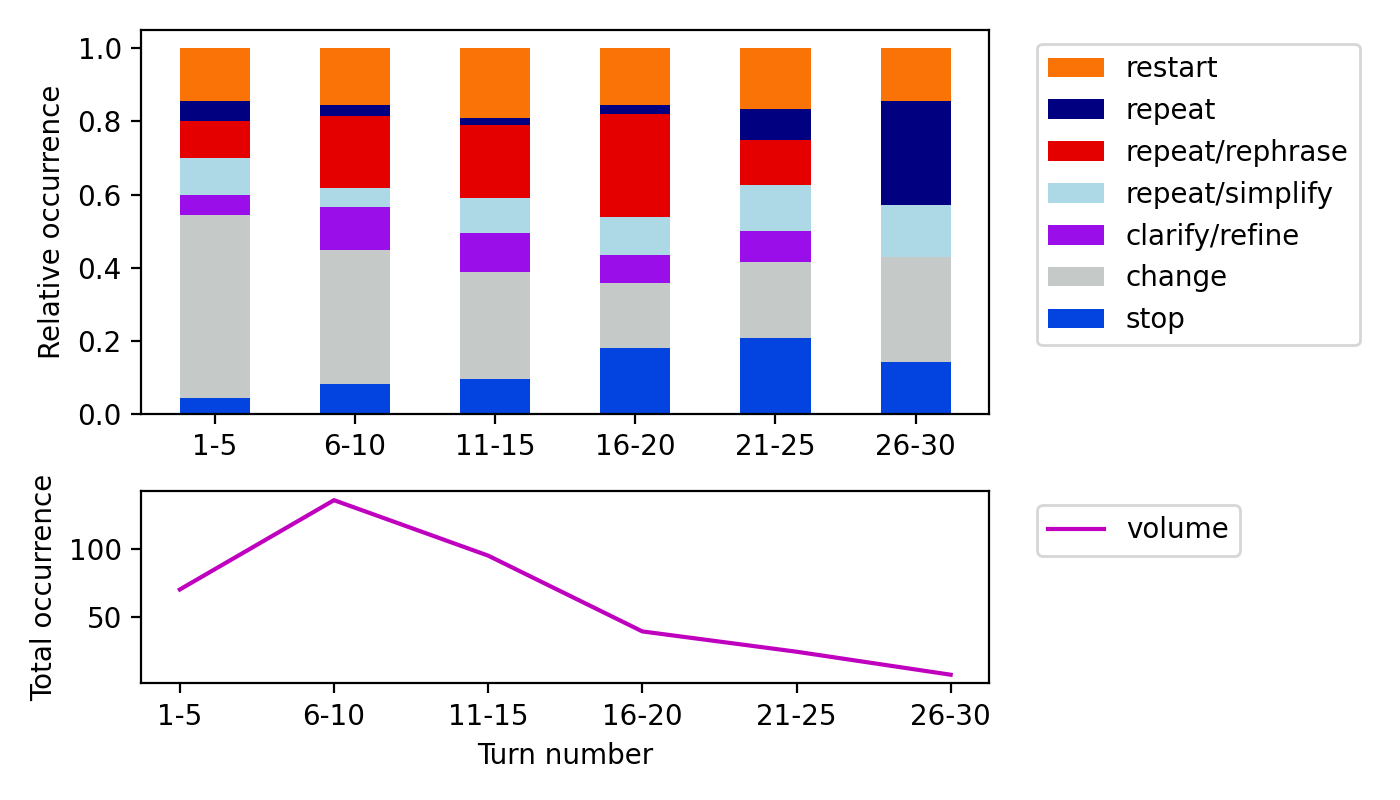} 
   \vspace{-1.5\baselineskip}
   \caption{Reformulation types over dialogue turns.}
    \label{fig:type_dist}
\end{figure}

Additionally, we ask: How do the type of reformulations change during the course of dialogues? 
In Fig.~\ref{fig:type_dist}, we show the distribution of various reformulation types across different stages (binned by turns of 5) on the bar plot (top), while the line plot (bottom) shows the absolute volume (i.e., total number of turns where reformulations occurred).
We observe, again, that rephrasing tends to happen earlier in dialogues than simplification.



\section{Simulating Utterance Reformulations}
\label{sec:sur}

In Section~\ref{sec:aur}, we have analyzed user behavior in terms of utterance reformulations.
Next, we address the question: \emph{How to incorporate human-like reformulation behavior into a state-of-the-art user simulator?}
Specifically, we propose the reformulation sequence generation task (cf. Sect~\ref{sec:sur:rsg}) to output a reformulation sequence based on a given utterance and reformulation type (rephrase or simplify), using type-guided text generation transformers (Sect.~\ref{sec:sur:rtu}). 
To ensure that the reformulations indeed achieve their intended goal, we further investigate ways to estimate the reading difficulty of the generated utterances (Sect.~\ref{sec:sur:eur}).  

\subsection{Reformulation Sequence Generation}
\label{sec:sur:rsg}

Users might attempt to reformulate the utterance when their requests are not understood. We describe these sequences of utterances resulted from reformulations as a reformulation sequence, \emph{$S^R$}. Each \emph{$S^R$} includes at least two user utterances (conversation turns) with the same user intent.
To simulate a reformulation sequence, we propose the reformulation sequence generation (\emph{RSG}) task. 
Given a user utterance \emph{$u_0$}, it aims to generate a sequence of reformations:
%
\begin{equation}
    RSG(u_0) = S^R =\{u_1, ..., u_n\},
    \label{eq:rsf}
\end{equation}
where $u$ denotes a user utterance.
For example, if $u_0$ is \emph{``I am into a movie like a tale of two cities''}, the user might \emph{rephrase} it to $u_1$ \emph{``movies similar to a tale of two cities''} when the system fails to understand the preference disclosure.
If failure continues, the user might \emph{repeat} $u_1$ in $u_2$ as \emph{``movies similar to a tale of two cities''} or even \emph{simplify} $u_2$ to $u_3$ \emph{``a tale of two cities''}.
Figure~\ref{fig:seq+smooth} illustrates the architecture we devise for this task.
As a result, the utterances in Eq.~\ref{eq:rsf} change sequentially and form a sequence of reformulation types:
\begin{equation}
    t = \{t_1, t_2, ... t_n\},
\end{equation}
where $t_n$ denotes the reformulation type from $u_{n-1}$ to $u_{n}$.
To model these transitions, we leverage the reformulation types identified in Table~\ref{tbl:reformulate}, and 
leverage the transition matrix $M$ in Section~\ref{sec:aur} to determine the reformulation type $t$.
%
Specifically, we randomly select the first reformulation type $t_1$ by the uniform distribution, and we further use it as a condition to generate the first reformulated utterance $u_1$.
E.g., given $u_0$ \emph{``Something light-hearted,''}
we select \emph{refine} as the first reformulation type $t_1$, and further use it as a condition to generate $u_1$ \emph{``I want a funny comedy that is light hearted.''} 
After the first generation, we update the distribution of reformulation type (``Dist update'' in Fig.~\ref{fig:seq+smooth}) by following the transition probability: 
\begin{equation}
\Vec{z}_{r}= \Vec{z}_r|\Vec{z}_{r-1} = M \cdot \Vec{z}_{r-1} ~, \label{update_distri}
\end{equation}
where $M$ represents the transition matrix, and the vector $\Vec{z}_r$ represents the probability distribution of reformulation types at step $r$. We update the distribution by calculating the dot product of the matrix and previous distribution vector.
After getting a new distribution of reformulation types, we select $t_2$ and use it as a condition to generate $u_2$. 
We then repeat this procedure in Fig.~\ref{fig:seq+smooth} to generate a sequence of reformulated utterances as output. 
We consider the sequence as a set of candidates which can be used to respond to the agent in simulation, if it fails to give a satisfactory reply.

\subsection{Type-guided Reformulation Utterance Generation}
\label{sec:sur:rtu}
In Section~\ref{sec:sur:rsg}, we formulated our task as a conditional sequence to sequence task. 
In each iteration, given $u_0$ and next reformulation type (which might be \emph{repeat}, \emph{rephrase}, \emph{simplify}, \emph{repeat}, or \emph{restart}), the model generates the next reformulated utterance via type-guided reformulation generation methods. 

We consider different transformer models, both a pre-trained language model and sequence-to-sequence models for addressing this task.
To train a language model, we concatenate the reformulation type, utterance, and reformulated utterance and further use masks to identify them.  We consider \emph{GPT2}~\citep{Radford:2019:GPT2} and use top-k sampling for the decoding algorithm, with $k$ set to 0.5. 
To train a sequence-to-sequence model, we consider both the utterance and the reformulation type $t$ as input, further concatenate them (separated by a special \texttt{</s>} token), and tokenize them as an input tensor. For the output, we tokenize the reformulated utterance as an output tensor. The sequence-to-sequence models we consider for comparison are \emph{BERT-to-BERT}~\citep{Jacob:2018:BERT} (denoted as \emph{BERT}), \emph{BART}~\citep{Mike:2019:BART}, and \emph{Text-to-Text Transfer Transformer (T5)}~\citep{Colin:2019:T5}.

\subsection{Filtering based on Reading Difficulty}
\label{sec:sur:eur}
One of the objectives of reformulation is to lower the reading difficulty for the generated utterance.
To avoid generating less understandable or linguistically more complex utterances than the original one, we include a module for estimating the reading difficulty in our simulation.
It classifies each generated utterance as acceptable or not.
Only the acceptable ones will be considered as candidates, i.e., the generated utterances with the $\surd$ label in Fig.~\ref{fig:seq+smooth}.

There has been a long history of research on language reading difficulty~\citep{Collins:2004:ALM, Wan:2018:ATS} and neural language understanding models advance the development recently~\citep{Jacob:2018:BERT, Alex:2018:NNAJ}.
As the first step towards this investigation in user simulation, we leverage a BERT-based classifier~\citep{Jacob:2018:BERT} fine-tuned on the CoLA dataset~\citep{Alex:2018:NNAJ}. This dataset is annotated based on whether the language is linguistically acceptable or not. Three types of sentences, which are \emph{morphological anomalies}, \emph{syntactic anomalies}, and \emph{semantic anomalies} are labeled as unacceptable.
Taking an example for our task, the binary classifier will label the utterance \emph{``I want to watch a good rating''} as unacceptable with unavailable meanings since it is considered as a morphological anomaly. 
To train the BERT classifier, we follow the exact same experimental setting as in~\citep{Jacob:2018:BERT}.

%
\begin{table*}[t]
    \footnotesize
	\caption{Automatic evaluation results. We test significance on the following pairs of methods: with type-guided vs. without, T5 vs. BART, BART vs. BERT, and BERT vs. GPT2. $\ddagger$ denotes significance at the 0.005 level.}
	\captionshrink
	\begin{tabular}{lllllllllllllll}
	\toprule
  & \multicolumn{4}{c}{\textbf{Movie}} &&     \multicolumn{4}{c}{\textbf{Travel}} && \multicolumn{4}{c}{\textbf{Hybrid}} \\
  \cline{2-5} \cline{7-10} \cline{12-15} 
  \textbf{Method} & \textbf{Rouge-1} & \textbf{Rouge-2} & \textbf{Rouge-L} & \textbf{BLEU} &&
  \textbf{Rouge-1} & \textbf{Rouge-2} & \textbf{Rouge-L} & \textbf{BLEU}&& 
  \textbf{Rouge-1} & \textbf{Rouge-2} & \textbf{Rouge-L} & \textbf{BLEU}\\
  \midrule

\textbf{GPT2} & 0.112 & 0.031 & 0.102 & 0.058 &&
         0.056 & 0.017 & 0.061 & 0.049 &&
         0.118 & 0.005 & 0.092 & 0.049\\
\textbf{GPT2-T}  & $0.113^\ddagger$ & $0.031^\ddagger$ & $0.106^\ddagger$ & $0.072^\ddagger$ &&
          $0.053^\ddagger$ & $0.015^\ddagger$ & $0.065^\ddagger$ & $0.072^\ddagger$ &&
          $0.125^\ddagger$ & $0.028^\ddagger$ & $0.137^\ddagger$ & $0.055^\ddagger$\\
\textbf{GPT2-R} & 0.072 & 0.021 & 0.096 & 0.037 &&
       0.056 & 0.015 & 0.102 & 0.042 &&
       0.106 & 0.017 & 0.124 & 0.105\\
\textbf{GPT2-TR} & \textbf{0.086$^\ddagger$} & \textbf{0.032$^\ddagger$} & \textbf{0.132$^\ddagger$} & \textbf{0.128$^\ddagger$} &&
       \textbf{0.152$^\ddagger$} & \textbf{0.031$^\ddagger$} & \textbf{0.127$^\ddagger$} & \textbf{0.073$^\ddagger$} &&
       \textbf{0.182$^\ddagger$} & \textbf{0.032$^\ddagger$} & \textbf{0.173$^\ddagger$} & \textbf{0.156$^\ddagger$} \\
\midrule      
\textbf{BERT} & 0.228 & 0.06 & 0.221 & 0.113 &&
         0.110 & 0.023 & 0.121 & 0.092 &&
         0.260 & 0.010 & 0.192 & 0.112\\
\textbf{BERT-T}  & $0.220^\ddagger$ & $0.069^\ddagger$ & $0.213^\ddagger$ & $0.121^\ddagger$ &&
          $0.121^\ddagger$ & $0.032^\ddagger$ & $0.132^\ddagger$ & $0.141^\ddagger$ &&
          $0.265^\ddagger$ & $0.054^\ddagger$ & $0.262^\ddagger$ & $0.109^\ddagger$\\
\textbf{BERT-R} & 0.223 & 0.09 & 0.193 & 0.132 &&
       0.132 & 0.038 & 0.231 & 0.086 &&
       0.243 & 0.032 & 0.234 & 0.213\\
\textbf{BERT-TR} & \textbf{0.324$^\ddagger$} & \textbf{0.092$^\ddagger$} & \textbf{0.312$^\ddagger$} & \textbf{0.132$^\ddagger$} &&
       \textbf{0.213$^\ddagger$} & \textbf{0.065$^\ddagger$} & \textbf{0.231$^\ddagger$} & \textbf{0.153$^\ddagger$} &&
       \textbf{0.321$^\ddagger$} & \textbf{0.072$^\ddagger$} & \textbf{0.342$^\ddagger$} & \textbf{0.327$^\ddagger$} \\
       
\midrule

\textbf{BART} & 0.281 & 0.108 & 0.279 & 2.229 &&
       0.281 & 0.094 & 0.281 & 1.572 &&
       0.301 & 0.120 & 0.302 & 2.562\\

\textbf{BART-T} & $0.504^\ddagger$ & $0.123^\ddagger$ & $0.501^\ddagger$ & $5.308^\ddagger$ &&
       $0.422^\ddagger$ & $0.102^\ddagger$ & $0.419^\ddagger$ & $5.286^\ddagger$ &&
       $0.519^\ddagger$ & $0.175^\ddagger$ & $0.516^\ddagger$ & $8.214^\ddagger$ \\
       
\textbf{BART-R}  & 0.343 & \textbf{0.154} & 0.328 & 3.326 &&
          0.356 & \textbf{0.158} & 0.312 & 2.359 &&
          0.362 & 0.163 & 0.396 & 3.382\\ 
          
\textbf{BART-TR}  & \textbf{0.521$^\ddagger$} & $0.124^\ddagger$ & \textbf{0.517$^\ddagger$} & \textbf{6.532$^\ddagger$} &&
         \textbf{0.437$^\ddagger$} & $0.100^\ddagger$ & \textbf{0.433$^\ddagger$} & \textbf{6.473$^\ddagger$} &&
         \textbf{0.585$^\ddagger$} & \textbf{0.285$^\ddagger$} & \textbf{0.532$^\ddagger$} & \textbf{9.481$^\ddagger$} \\

  \midrule
\textbf{T5} & 0.410 & 0.243 & 0.410& 3.472&&
0.369& 0.262 & 0.367& 2.740 && 
0.400 & 0.245 & 0.399& 4.739 \\
\textbf{T5-T} & $0.742^\ddagger$ & $0.320^\ddagger$ & $0.739^\ddagger$ & $19.031^\ddagger$ &&
$0.741^\ddagger$ & \textbf{0.297$^\ddagger$} & $0.738^\ddagger$ & $21.485^\ddagger$&& 
$0.783^\ddagger$ & $0.340^\ddagger$ & $0.782^\ddagger$ & $20.987^\ddagger$ \\

\textbf{T5-R} & 0.401 & 0.261 & 0.399 & 3.578 &&
       0.365 & 0.240 & 0.340 & 3.253&&
       0.420 & 0.270 & 0.430 & 6.242\\
       
\textbf{T5-TR} & \textbf{0.766$^\ddagger$} & \textbf{0.321$^\ddagger$} & \textbf{0.740$^\ddagger$} & \textbf{20.739$^\ddagger$} &&
        \textbf{0.755$^\ddagger$} & $0.294^\ddagger$ & \textbf{0.753$^\ddagger$} & \textbf{23.242$^\ddagger$} &&
         \textbf{0.792$^\ddagger$} & \textbf{0.350$^\ddagger$} & \textbf{0.785$^\ddagger$} & \textbf{22.487$^\ddagger$} \\
	\bottomrule
	\end{tabular}
\label{tbl:re1}
\end{table*}

\section{Experimental Setup}
\label{e}

\subsection{Constructing Training Collection}
\label{e:ctc}

\noindent
Recall that we have collected data from five conversational agents on three domains.
We construct training data for the utterance generation task by 
taking these logged dialogues from the user study and further enrich them via another crowdsourcing task.
%
\subsubsection{Logged Dialogues}
\label{sec:ctc:ld}
To integrate the logged dialogues from the user study, we consider user utterances labeled with a \emph{reformulation} type (cf. Sect.~\ref{sec:aur:ra}) and further back-trace them to the previous user utterance as the \emph{original utterance}. 
The original utterance, reformulation type, and reformulated utterance form a triad $\langle u_0, t, u^r_t \rangle$.
We consider the triad records from all the agents except \emph{Music}, since there are only 5 reformulation sequences in all 30 logged dialogues for the Music domain. For the other 4 agents, we combine the 30 dialogues per agent as our initial dataset.
\subsubsection{Crowdsourcing}
To enrich the diversity of the initial dataset, we conduct an utterance rewriting crowdsourcing task on Amazon MTurk.
We present workers with a scenario that a conversational agent fails to understand the request when recommending movies or travel plans. 
We instruct the annotators to rewrite an utterance by reformulating, refining, or simplifying it. 
We sample 208 triads from Sect.~\ref{sec:ctc:ld} when the system fails to understand the requests. 
The original reformulation sequence serves as the input to simulate the scenario and enable each rewriting task. 
We invite nine annotators per utterance to rewrite it based on the intent, and harvest about 1.8k rewritten utterances by 265 workers. 
These collected utterances are integrated with the logged training data by domain, intent, and reformulation type.

\subsection{Experimental Methods}

\subsubsection{Implementation Details}

We use the Huggingface~\citep{Thomas:2019:HT} implementations of BERT-to-BERT, Bart, T5, and GPT2. For the BERT-to-BERT model, we use the Encoder-Decoder Models frame with both initialized with the ``bert-uncased'' model. To finetune Bart, T5, and GPT2, we use the Conditional Generation model on Huggingface, and ``Facebook/bart-large,'' ``t5,'' and ``gpt2'' pre-trained models.

To examine the generalization capabilities of these models, we leverage the training data in three ways: only \emph{movie} domain, only \emph{travel} domain, and \emph{hybrid}. The \emph{hybrid} mode is meant to represent a domain-independent approach by combining the \emph{travel} and \emph{movie} datasets.
We only consider \emph{refine}, \emph{rephrase}, \emph{simplify}, \emph{repeat}, and \emph{restart} as the reformulation types in the experiments, as \emph{stop} and \emph{change} would change the intent of original utterances. We leave the reformulation task with changed intent for future work.
Each original utterance $u_0$ has more than one reformulated candidates under the same reformulation type.  
We parallelize the records and feed them into the model in the triad data format as mentioned in Sect.~\ref{e:ctc}. 
The training process is through 5-fold cross-validation. 
We set the batch size as 10 and initialize the model with default parameters.
To test the model, we follow the sequence generation procedure in Fig.~\ref{fig:seq+smooth}. 
We set the sequence length as 3, which is the average length of reformulation sequence in the logged dialogues.
While updating the reformulation type from the distribution, it will retry if \emph{change} or \emph{stop} is encountered. 

\subsubsection{Utterance Generation Variants}
We consider four variants of each utterance generation method. We use \textbf{Model} to represent any text generation model we select in Sect.~\ref{sec:sur:rtu}.
\begin{itemize}[leftmargin=0.5cm]
    \item \textbf{[Model]}: The vanilla ones are without reformulation type as input and reading difficulty filter (first row of each block in Table~\ref{tbl:re1}).
    \item \textbf{[Model]-T}: The second rows of each block in Table~\ref{tbl:re1} are the ones trained with guided reformulation type.
    \item \textbf{[Model]-R}: The third rows are the methods considering the component of reading difficulty filter.\footnote{When adding the reading difficulty filter on the BERT-to-BERT model, we relax the boundary of detecting the non-linguistic utterance, since the utterance generated from the BERT-to-BERT model mostly can't pass the linguistic test}
    \item \textbf{[Model]-TR}: The last rows consider both type and reading difficulty.
\end{itemize}


%

\section{Evaluation and Analysis}
\label{e:em}

We evaluate the approaches devised for our novel simulation task (reformulation sequence generation) using automatic evaluation (Sect.~\ref{e:em:ae}), expert human evaluation (Sect.~\ref{e:lhe}), and crowdsourcing human evaluation (Sect.~\ref{e:ihe}).


\begin{table*}[t]
	\caption{Side-by-side comparison results, with human evaluators guessing which of two dialogs within a given domain was performed by a simulated user (Win) vs. a real one (Loss); a Tie is given when the evaluator could not decide.}
	\captionshrink	
	\small
	\begin{tabular}{llllllllllll}
	\toprule
	  & & \textbf{Movie} &&&& \textbf{Travel} &&&& \textbf{All} \\  
	  \cline{2-4} \cline{6-8} \cline{10-12}
	 	  & \textbf{Win} & \textbf{Lose} & \textbf{Tie} &&\textbf{Win} & \textbf{Lose} & \textbf{Tie} && \textbf{Win} & \textbf{Lose} & \textbf{Tie}  \\  
	\midrule
	 \textbf{BART} & \textbf{39 (27\%)} & 53 (36\%) & 55 (37\%)
	              && \textbf{27 (33\%)} & 29 (36\%) & 25 (31\%) 
	             &&  66 (29\%) & 82 (36\%) & 80 (35\%) \\	
	 \textbf{BART-TR}  & 34 (23\%) & 35 (24\%) & 78 (53\%) 
	                  && 24 (30\%) & 25 (31\%) & 32 (39\%) 
	                  && 58 (25\%) & 60 (26\%) & 110 (49\%)  \\
	 \textbf{T5}  & 36 (24\%) & 53 (36\%) & 58 (40\%) 
	    && 16 (20\%) & 35 (43\%) & 30 (37\%) 
	    && 52 (22\%) & 88 (39\%) & 88 (39\%)  \\	
	 \textbf{T5-TR}  & \textbf{49 (33\%)} & 45 (31\%) & 53 (36\%) 
	            && \textbf{27 (33\%)} & 19 (23\%) & 35 (44\%) 
	            && \textbf{76 (33\%)} & 64 (28\%) & 88 (39\%) \\	
	\bottomrule
	\end{tabular}
\label{tbl:human}
\end{table*}

\begin{table}[t]
	\caption{Expert human evaluation results. The columns are \textbf{G}rammar, \textbf{D}ifficulty, and \textbf{T}ype. Lower D values correspond to less difficult text (which is desired).}
 	\captionshrink
    \footnotesize
	\begin{tabular}{p{1.38cm}p{ 0.35cm}p{ 0.35cm}p{ 0.35cm}p{ 0.35cm}p{ 0.35cm}p{ 0.35cm}p{ 0.35cm}p{ 0.35cm}p{ 0.35cm}}
	\toprule
	& \multicolumn{3}{c}{\textbf{Movie}} & \multicolumn{3}{c}{\textbf{Travel}} & \multicolumn{3}{c}{\textbf{Hybrid}} \\ 
	\cline{2-4} 	\cline{5-7} \cline{8-10}
	 &\textbf{G} & \textbf{D} & \textbf{T} &
	 \textbf{G} & \textbf{D} & \textbf{T} &
	 \textbf{G} & \textbf{D} & \textbf{T} \\
	\midrule
	\textbf{BART}& 
		0.90 & 1.02 & 0.14 &
		\textbf{1.00} & 1.00 & 0.15  &
		0.94 & 1.00 & 0.14\\
	\textbf{BART-T} & 
		0.98 & 0.97 & 0.48 &
		0.97 & 0.86 & 0.41 &
		0.98 & 0.83 & 0.55\\
	\textbf{BART-R} & 
		\textbf{1.00} & 0.87 & 0.23 &
		\textbf{1.00} & 0.90 & 0.18 &
		0.98 & 0.88 & 0.22 \\
	\textbf{BART-TR} & 
		\textbf{1.00} & \textbf{0.75} & \textbf{0.65} &
		0.98 & \textbf{0.68} & \textbf{0.64} &
		\textbf{1.00} & \textbf{0.73} & \textbf{0.72} \\
	\midrule
	\textbf{T5} & 
		\textbf{1.000} & 0.95 & 0.22 &
		\textbf{1.00} & 0.95 & 0.14 &
		0.96 & 0.93 & 0.16 \\
	\textbf{T5-T} & 
		0.97 & 0.87 & 0.72 &
		\textbf{1.00} & 0.81 & 0.82 &
		0.98 & 0.88 & 0.72\\
	\textbf{T5-R} & 
		\textbf{1.00} & 0.82 & 0.34 & 
		\textbf{1.00} & 0.85 & 0.25 &
		0.98 & 0.81 & 0.27 \\
	\textbf{T5-TR} & 
		\textbf{0.98} & \textbf{0.69} & \textbf{0.78} &
		\textbf{1.00} & \textbf{0.73} & \textbf{0.88} &
		\textbf{0.98} & \textbf{0.72} & \textbf{0.82} \\
	\bottomrule
	\end{tabular}
\label{tbl:human_lab}
\end{table}
%

\subsection{Automatic Evaluation with NLP Metrics}
\label{e:em:ae}

For automatic evaluation we use ROUGE \citep{Lin:2004:ROUGE} and BLEU \citep{Papineni:2002:BLEU} as our evaluation metrics. Both metrics are often used on text summarization tasks. 
To guarantee the stability of experimental results, we generate the reformulation sequence five times.
We further filter the generated sequences via the utterance reading difficulty component to exclude hard-to-read utterances. 
We calculate the metrics using micro-averaging and further average the score across each sequence. 
All the results are reported in Table~\ref{tbl:re1}. 
When evaluating the effect of type guiding ([Method]-T), we find that most of the type-guided methods outperform the vanilla methods (without type-guiding) significantly ([Method]).  
We also note the performance of methods without type-guiding is not robust (it can vary a lot). 
Among the three domains, the type-guided methods for hybrid perform comparably with the single domain method, indicates the generalization capabilities of the combined model.

Filtering based on utterance reading difficulty ([Method]-R) can improve performance in some cases. However, it can enhance performance significantly and consistently when combined with type guidance. Indeed, the [Method]-TR achieves the best performance in almost all cases. Therefore, we conclude that type-guiding and reading difficulty estimation often help on their own, but not consistently across all methods and domains. However, combining the two always works. The improvements are substantial and significant. 

In terms of transformer selection, we find that the sequence-to-sequence models significantly outperform the language model (GPT2) on all measures. 
Amongst the pre-trained sequence-to-sequence models, T5 performs the best.

%

\subsection{Expert Human Evaluation}
\label{e:lhe}

Next, we employ human experts (two of the authors) to evaluate the syntax and semantic quality of the generated reformulations. Specifically, we evaluate three aspects: \emph{Grammar}, \emph{Reading Difficulty}, and \emph{Type Accuracy}. 
For \emph{movie} and \emph{hybrid}, we randomly sample 200 reformulation sequences; for \emph{travel} we use all 150 reformulation sequences. 
We only evaluate the BART and T5 models as the majority of the utterances generated by the BERT-to-BERT models fail to pass our reading difficulty filter.
The annotations are agreed by the authors and the results are reported in Table~\ref{tbl:human_lab}.

Our observations are as follows.
The \emph{grammar correctness} scores are very high (above 0.95) for all domains and methods, demonstrating that that these neural models are capable of generating sentences free of grammatical mistakes. 
\emph{Reading difficulty} estimates whether the generated utterances are more difficult to understand (2), have the same reading difficulty (1), or easier to understand (0) than the original utterance. That is, it ranges from 0 to 2, and lower values are better.
We can see the scores are decreasing from the vanilla methods to ones with both type-guiding and reading difficulty filter. 
Thus, the reading difficulty filter indeed contributes achieving the task objectives.
Finally, \emph{type accuracy} measures whether the generated reformulation is consistent with the target reformulation type or not. 
We can see that the type-guided methods perform better in this regard. 

Comparing the results with those from automatic evaluation, we find that the same observations hold regarding the improvements brought in by type-guiding and reading difficulty filtering, as well as generalizability (hybrid vs. single-domain performance).



\subsection{Crowdsourcing Human Evaluation}
\label{e:ihe}

To investigate the end-to-end performance of our model, we further evaluate the simulated utterances via crowdsourcing to test whether the generated sequence is indistinguishable from conversations performed by real users, following~\citep{Zhang:2020:ECR}.
We sample 46 dialogues from the logged data, and replace the reformulation utterances in the original dialogues with our simulated utterances according to the same reformulation type. For example, in Table~\ref{tbl:cht}, the reformulations happens 2 times in the original dialogue (colored in blue and red). We replace these original reformulations using our generated utterances.
We randomly generate a total of 302 dialogue pieces. We place the original and simulated dialogue pieces side-by-side, in a random order, and present them to 3 crowd workers on Amazon MTurk.
They are asked to choose which of the two dialogues is performed purely by a human; they are specifically instructed to compare the highlighted reformulated sequences. A tie is permitted when it is hard to distinguish.
Additionally, workers are requested to give a brief explanation behind their choice. Options without explanations are filtered out. 
The results are presented in Table \ref{tbl:human}. 
We can see that the improvements of considering type-guided and reading difficulty filter are consistent with automatic and expert human evaluation. At the same time, the performance gaps are smaller.
It is worth noting that 40\% of pairs are not distinguishable, which also attests to the human-likeness of simulated responses.
\subsection{Analysis}
Next, we analyze the explanations given by crowd workers in Section~\ref{e:ihe} for their choice of which of the dialogues was performed purely by a human.
%
We observe the same groups of reasons with~\citep{Zhang:2020:ECR} based on style and content. 
\begin{table}[t]
\small
    \centering
        \caption{An example reformulation dialogue piece and its corresponding simulated dialogue piece.}
    \label{tbl:cht}
    \captionshrink
    \begin{tabular}{l}
    \toprule
    \textbf{Original Reformulation Piece} \\
    \midrule
    \textbf{User}: OK, now I want to know more about restaurants in Dubai \\
    \textbf{Agent}: Hotel? Apartment? or a bunk bed? \\
    \textbf{User}: \textcolor{blue}{Need to know about restaurants} \\
    \textbf{Agent}: Sorry, I didn't get that. Can you rephrase? \\
    \textbf{User}: \textcolor{red}{Restaurant}  \\
    \midrule
    \textbf{Simulated Reformulation Piece} \\
    \midrule
    \textbf{User}: OK, now I want to know more about restaurants in Dubai \\
    \textbf{Agent}: Hotel? Apartment? or a bunk bed? \\
    \textbf{User}: \textcolor{blue}{Can you find me a restaurant in Dubai?} \\
    \textbf{Agent}: Sorry, I didn't get that. Can you rephrase? \\
    \textbf{User}: \textcolor{red}{Places for dinner} \\
    \bottomrule
    \end{tabular}
    \vspace*{-0.5\baselineskip}
\end{table}

For example, the aspects mentioned in terms of style include ``\emph{this conversation seemed more natural and realistic to human conversation}'' (Realisticity),  ``\emph{the user keep trying to restart to get the right answer}'' (Engagement), and ``\emph{this user seems to get a bit frustrated with the bot and adapts to it}'' (Emotion).
Those based on content are include ``\emph{the first one seems to understand the request a bit better}'' (Response),
``\emph{the top is likely the human as there are punctuation and varied capitalizations}'' (Grammar), and 
``\emph{it says a lot of words that seem robotic}'' (Length).

There is also a new group of reasons focusing on the effect of reformulation.
Most annotators take reformulation as being positive, e.g., ``\emph{they are asked to clarify their question, then they rephrase it so the Bot Agent can understand}.'' 
But there are also negative cases, e.g., 
``\emph{the user in 1 appears to get stuck in a loop and ends up requesting nonsensical topics}.''
To summarize, integrating reformulation behavior contributes to making the simulation more human-like. However, low-quality reformulations, like sticking to nonsensical topics, might hurt.


\vspace{0.5\baselineskip}
\section{Conclusions}

This work has focused on one particular user trait, utterance reformation in response to failed replies, to create more human-like user simulation. 
We have started our investigation with a user study, where interactions with five conversational agents across three domains were requested. 
The main outcome of the user study are seven common reformulation types that occur in conversational recommendations.
To incorporate the findings from the user study into user simulation approaches, we have proposed a reformulation sequence generation task based on a given utterance.
We have developed methods using type-guided text generation transformers to address this task, and have also included a way to estimate reading difficulty.
The proposed methods have been compared using automatic evaluation, expert human evaluation, and crowdsourcing human evaluation. 
We have found that type-guiding or reading difficulty filtering alone does not always guarantee improvements, but their combination has been shown to improve performance significantly and substantially across methods, domains, and evaluation metrics.  Furthermore, experiments on a hybrid-domain dataset have shown the generalizability of our method.
The utterance reformulation dataset we have developed as part of this study consists of 1.8k reformulated utterances for 150 dialogues, which can be a useful resource for training more realistic simulators.

\textbf{\emph{Limitations.}}
Our user study has focused on modeling the reformulation behavior of a persistent user. 
In practice, a simulator would need to model users with different personality types, ranging from non-persistent to maximally persistent. Estimating the distribution of persistence for the simulated user population requires further investigation.
We also note that our model is trained on a particular set of agents and might suffer from generalizability; for new agents, data may need to be collected for fine-tuning.

\textbf{\emph{Future work.}}
We see several avenues for extending our work in the future. 
First, the improved, more human-like reformulation behavior is to be incorporated into a larger simulation framework, like~\citep{Zhang:2020:ECR}, and evaluated in the context of an end-to-end task. 
There are also additional dimensions of reformulation behavior to consider, such as the users' knowledge and preferences, changing intents, style consistency of generated reformulations, and effects of the success rate of previous dialogue.
\begin{acks}
This work was partially supported by the National Natural Science Foundation of China (61973272).
\end{acks}

\bibliographystyle{ACM-Reference-Format}
\balance
\bibliography{nlu}


\begin{thebibliography}{53}


\ifx \showCODEN    \undefined \def \showCODEN     #1{\unskip}     \fi
\ifx \showDOI      \undefined \def \showDOI       #1{#1}\fi
\ifx \showISBNx    \undefined \def \showISBNx     #1{\unskip}     \fi
\ifx \showISBNxiii \undefined \def \showISBNxiii  #1{\unskip}     \fi
\ifx \showISSN     \undefined \def \showISSN      #1{\unskip}     \fi
\ifx \showLCCN     \undefined \def \showLCCN      #1{\unskip}     \fi
\ifx \shownote     \undefined \def \shownote      #1{#1}          \fi
\ifx \showarticletitle \undefined \def \showarticletitle #1{#1}   \fi
\ifx \showURL      \undefined \def \showURL       {\relax}        \fi
\providecommand\bibfield[2]{#2}
\providecommand\bibinfo[2]{#2}
\providecommand\natexlab[1]{#1}
\providecommand\showeprint[2][]{arXiv:#2}

\bibitem[\protect\citeauthoryear{Aliannejadi, Zamani, Crestani, and
  Croft}{Aliannejadi et~al\mbox{.}}{2019}]%
        {Aliannejadi:2019:ACQ}
\bibfield{author}{\bibinfo{person}{Mohammad Aliannejadi},
  \bibinfo{person}{Hamed Zamani}, \bibinfo{person}{Fabio Crestani}, {and}
  \bibinfo{person}{W.~Bruce Croft}.} \bibinfo{year}{2019}\natexlab{}.
\newblock \showarticletitle{Asking Clarifying Questions in Open-Domain
  Information-Seeking Conversations}. In \bibinfo{booktitle}{\emph{Proceedings
  of the 42nd International ACM SIGIR Conference on Research and Development in
  Information Retrieval}} \emph{(\bibinfo{series}{SIGIR '19})}.
  \bibinfo{pages}{475--484}.
\newblock


\bibitem[\protect\citeauthoryear{Asri, He, and Suleman}{Asri
  et~al\mbox{.}}{2016}]%
        {AsriHS:2016:ASM}
\bibfield{author}{\bibinfo{person}{Layla~El Asri}, \bibinfo{person}{Jing He},
  {and} \bibinfo{person}{Kaheer Suleman}.} \bibinfo{year}{2016}\natexlab{}.
\newblock \showarticletitle{A Sequence-to-Sequence Model for User Simulation in
  Spoken Dialogue Systems}. In \bibinfo{booktitle}{\emph{Proceedings of the
  17th Annual Conference of the International Speech Communication
  Association}} \emph{(\bibinfo{series}{Interspeech '16})}.
  \bibinfo{pages}{1151--1155}.
\newblock


\bibitem[\protect\citeauthoryear{Azzopardi, Dubiel, Halvey, and
  Dalton}{Azzopardi et~al\mbox{.}}{2018}]%
        {Leif:2018:CAI}
\bibfield{author}{\bibinfo{person}{Leif Azzopardi}, \bibinfo{person}{Mateusz
  Dubiel}, \bibinfo{person}{Martin Halvey}, {and} \bibinfo{person}{Jeffery
  Dalton}.} \bibinfo{year}{2018}\natexlab{}.
\newblock \showarticletitle{Conceptualizing Agent-human Interactions during the
  Conversational Search Process}. In \bibinfo{booktitle}{\emph{The Second
  International Workshop on Conversational Approaches to Information
  Retrieval}} \emph{(\bibinfo{series}{CAIR '18})}.
\newblock


\bibitem[\protect\citeauthoryear{Balog}{Balog}{2021}]%
        {Balog:2021:DESIRES}
\bibfield{author}{\bibinfo{person}{Krisztian Balog}.}
  \bibinfo{year}{2021}\natexlab{}.
\newblock \showarticletitle{Conversational {AI} from an Information Retrieval
  Perspective: {R}emaining Challenges and a Case for User Simulation}. In
  \bibinfo{booktitle}{\emph{Proceedings of the 2nd International Conference on
  Design of Experimental Search \& Information REtrieval Systems}}
  \emph{(\bibinfo{series}{DESIRES '21})}. \bibinfo{pages}{80--90}.
\newblock


\bibitem[\protect\citeauthoryear{Balog, Maxwell, Thomas, and Zhang}{Balog
  et~al\mbox{.}}{2021a}]%
        {Balog:2021:SIGIRForum}
\bibfield{author}{\bibinfo{person}{Krisztian Balog}, \bibinfo{person}{David
  Maxwell}, \bibinfo{person}{Paul Thomas}, {and} \bibinfo{person}{Shuo Zhang}.}
  \bibinfo{year}{2021}\natexlab{a}.
\newblock \showarticletitle{Report on the 1st Simulation for Information
  Retrieval Workshop (Sim4IR 2021) at SIGIR 2021}.
\newblock \bibinfo{journal}{\emph{SIGIR Forum}} \bibinfo{volume}{52},
  \bibinfo{number}{2} (\bibinfo{date}{Dec.} \bibinfo{year}{2021}),
  \bibinfo{pages}{11--26}.
\newblock


\bibitem[\protect\citeauthoryear{Balog, Radlinski, and Karatzoglou}{Balog
  et~al\mbox{.}}{2021b}]%
        {Balog:2021:OIM}
\bibfield{author}{\bibinfo{person}{Krisztian Balog}, \bibinfo{person}{Filip
  Radlinski}, {and} \bibinfo{person}{Alexandros Karatzoglou}.}
  \bibinfo{year}{2021}\natexlab{b}.
\newblock \showarticletitle{On Interpretation and Measurement of Soft
  Attributes for Recommendation}. In \bibinfo{booktitle}{\emph{Proceedings of
  the 44th International ACM SIGIR Conference on Research and Development in
  Information Retrieval}} \emph{(\bibinfo{series}{SIGIR '21})}.
  \bibinfo{pages}{890--899}.
\newblock


\bibitem[\protect\citeauthoryear{Belz and Reiter}{Belz and Reiter}{2006}]%
        {Belz:2006:CAA}
\bibfield{author}{\bibinfo{person}{Anja Belz} {and} \bibinfo{person}{Ehud
  Reiter}.} \bibinfo{year}{2006}\natexlab{}.
\newblock \showarticletitle{Comparing Automatic and Human Evaluation of NLG
  Systems}. In \bibinfo{booktitle}{\emph{Proceedings of the 11th Conference of
  the European Chapter of the Association for Computational Linguistics}}
  \emph{(\bibinfo{series}{EACL '06})}. \bibinfo{pages}{313--320}.
\newblock


\bibitem[\protect\citeauthoryear{Brown and B.Forsythe}{Brown and
  B.Forsythe}{1974}]%
        {Morton:1974:RTE}
\bibfield{author}{\bibinfo{person}{Morton~B. Brown} {and} \bibinfo{person}{Alan
  B.Forsythe}.} \bibinfo{year}{1974}\natexlab{}.
\newblock \showarticletitle{Robust Tests for the Equality of Variances}.
\newblock \bibinfo{journal}{\emph{J. Amer. Statist. Assoc.}}
  \bibinfo{volume}{69}, \bibinfo{number}{346} (\bibinfo{year}{1974}),
  \bibinfo{pages}{364--367}.
\newblock


\bibitem[\protect\citeauthoryear{Ca{\~n}amares, Castells, and
  Moffat}{Ca{\~n}amares et~al\mbox{.}}{2020}]%
        {Canameres:2020:OEO}
\bibfield{author}{\bibinfo{person}{Roc{\'\i}o Ca{\~n}amares},
  \bibinfo{person}{Pablo Castells}, {and} \bibinfo{person}{Alistair Moffat}.}
  \bibinfo{year}{2020}\natexlab{}.
\newblock \showarticletitle{Offline Evaluation Options for Recommender
  Systems}.
\newblock \bibinfo{journal}{\emph{Inf. Retr. J.}} \bibinfo{volume}{23},
  \bibinfo{number}{4} (\bibinfo{year}{2020}), \bibinfo{pages}{387--410}.
\newblock


\bibitem[\protect\citeauthoryear{Collins-Thompson and Callan}{Collins-Thompson
  and Callan}{2004}]%
        {Collins:2004:ALM}
\bibfield{author}{\bibinfo{person}{Kevyn Collins-Thompson} {and}
  \bibinfo{person}{James~P. Callan}.} \bibinfo{year}{2004}\natexlab{}.
\newblock \showarticletitle{A Language Modeling Approach to Predicting Reading
  Difficulty}. In \bibinfo{booktitle}{\emph{Proceedings of the Human Language
  Technology Conference of the North {A}merican Chapter of the Association for
  Computational Linguistics}} \emph{(\bibinfo{series}{HLT-NAACL '04})}.
  \bibinfo{pages}{193--200}.
\newblock


\bibitem[\protect\citeauthoryear{Dalton, Xiong, and Callan}{Dalton
  et~al\mbox{.}}{2020}]%
        {Dalton:2020:TCC}
\bibfield{author}{\bibinfo{person}{Jeffrey Dalton}, \bibinfo{person}{Chenyan
  Xiong}, {and} \bibinfo{person}{Jamie Callan}.}
  \bibinfo{year}{2020}\natexlab{}.
\newblock \bibinfo{title}{{TREC} CAsT 2019: The Conversational Assistance Track
  Overview}.
\newblock
\newblock
\showeprint[arxiv]{2003.13624}~[cs.IR]


\bibitem[\protect\citeauthoryear{Deriu, Rodrigo, Otegi, Echegoyen, Rosset,
  Agirre, and Cieliebak}{Deriu et~al\mbox{.}}{2019}]%
        {Jan:2020:SEM}
\bibfield{author}{\bibinfo{person}{Jan Deriu}, \bibinfo{person}{{\'{A}}lvaro
  Rodrigo}, \bibinfo{person}{Arantxa Otegi}, \bibinfo{person}{Guillermo
  Echegoyen}, \bibinfo{person}{Sophie Rosset}, \bibinfo{person}{Eneko Agirre},
  {and} \bibinfo{person}{Mark Cieliebak}.} \bibinfo{year}{2019}\natexlab{}.
\newblock \bibinfo{title}{Survey on Evaluation Methods for Dialogue Systems}.
\newblock
\newblock
\showeprint[arxiv]{1905.04071}~[cs.CL]


\bibitem[\protect\citeauthoryear{Devlin, Chang, Lee, and Toutanova}{Devlin
  et~al\mbox{.}}{2019}]%
        {Jacob:2018:BERT}
\bibfield{author}{\bibinfo{person}{Jacob Devlin}, \bibinfo{person}{Ming-Wei
  Chang}, \bibinfo{person}{Kenton Lee}, {and} \bibinfo{person}{Kristina
  Toutanova}.} \bibinfo{year}{2019}\natexlab{}.
\newblock \showarticletitle{{BERT: Pre-training of Deep Bidirectional
  Transformers for Language Understanding}}. In
  \bibinfo{booktitle}{\emph{Proceedings of the 2019 Conference of the North
  American Chapter of the Association for Computational Linguistics}}
  \emph{(\bibinfo{series}{ACL '19})}. \bibinfo{pages}{4171--4186}.
\newblock


\bibitem[\protect\citeauthoryear{Gao, Lei, He, de~Rijke, and Chua}{Gao
  et~al\mbox{.}}{2021}]%
        {Gao:2021:ACC}
\bibfield{author}{\bibinfo{person}{Chongming Gao}, \bibinfo{person}{Wenqiang
  Lei}, \bibinfo{person}{Xiangnan He}, \bibinfo{person}{Maarten de Rijke},
  {and} \bibinfo{person}{Tat-Seng Chua}.} \bibinfo{year}{2021}\natexlab{}.
\newblock \bibinfo{title}{Advances and Challenges in Conversational Recommender
  Systems: A Survey}.
\newblock
\newblock
\showeprint[arxiv]{2101.09459}~[cs.CL]


\bibitem[\protect\citeauthoryear{Gao, Galley, and Li}{Gao
  et~al\mbox{.}}{2018}]%
        {Gao:2018:NAC}
\bibfield{author}{\bibinfo{person}{Jianfeng Gao}, \bibinfo{person}{Michel
  Galley}, {and} \bibinfo{person}{Lihong Li}.} \bibinfo{year}{2018}\natexlab{}.
\newblock \showarticletitle{Neural Approaches to Conversational AI}. In
  \bibinfo{booktitle}{\emph{Proceedings of the 56th Annual Meeting of the
  Association for Computational Linguistics: Tutorial Abstracts}}
  \emph{(\bibinfo{series}{ACL '18})}. \bibinfo{pages}{2--7}.
\newblock


\bibitem[\protect\citeauthoryear{Griol, Carb\'{o}, and Molina}{Griol
  et~al\mbox{.}}{2013}]%
        {Griol:2013:ADS}
\bibfield{author}{\bibinfo{person}{David Griol}, \bibinfo{person}{Javier
  Carb\'{o}}, {and} \bibinfo{person}{Jos\'{e}~M. Molina}.}
  \bibinfo{year}{2013}\natexlab{}.
\newblock \showarticletitle{An Automatic Dialog Simulation Technique to Develop
  and Evaluate Interactive Conversational Agents}.
\newblock \bibinfo{journal}{\emph{Appl. Artif. Intell.}} \bibinfo{volume}{27},
  \bibinfo{number}{9} (\bibinfo{date}{oct} \bibinfo{year}{2013}),
  \bibinfo{pages}{759--780}.
\newblock


\bibitem[\protect\citeauthoryear{Habib, Zhang, and Balog}{Habib
  et~al\mbox{.}}{2020}]%
        {Habib:2020:IMA}
\bibfield{author}{\bibinfo{person}{Javeria Habib}, \bibinfo{person}{Shuo
  Zhang}, {and} \bibinfo{person}{Krisztian Balog}.}
  \bibinfo{year}{2020}\natexlab{}.
\newblock \showarticletitle{IAI MovieBot: A Conversational Movie Recommender
  System}. In \bibinfo{booktitle}{\emph{Proceedings of the 29th ACM
  International Conference on Information and Knowledge Management}}
  \emph{(\bibinfo{series}{CIKM '20})}. \bibinfo{pages}{3405--3408}.
\newblock


\bibitem[\protect\citeauthoryear{Huang, Oosterhuis, de~Rijke, and van
  Hoof}{Huang et~al\mbox{.}}{2020}]%
        {Huang:2020:KDB}
\bibfield{author}{\bibinfo{person}{Jin Huang}, \bibinfo{person}{Harrie
  Oosterhuis}, \bibinfo{person}{Maarten de Rijke}, {and} \bibinfo{person}{Herke
  van Hoof}.} \bibinfo{year}{2020}\natexlab{}.
\newblock \showarticletitle{Keeping Dataset Biases out of the Simulation: A
  Debiased Simulator for Reinforcement Learning Based Recommender Systems}. In
  \bibinfo{booktitle}{\emph{Fourteenth ACM Conference on Recommender Systems}}
  \emph{(\bibinfo{series}{RecSys '20})}. \bibinfo{pages}{190--199}.
\newblock


\bibitem[\protect\citeauthoryear{Jannach, Manzoor, Cai, and Chen}{Jannach
  et~al\mbox{.}}{2020}]%
        {Jannach:2020:SCR}
\bibfield{author}{\bibinfo{person}{Dietmar Jannach}, \bibinfo{person}{Ahtsham
  Manzoor}, \bibinfo{person}{Wanling Cai}, {and} \bibinfo{person}{Li Chen}.}
  \bibinfo{year}{2020}\natexlab{}.
\newblock \bibinfo{title}{A Survey on Conversational Recommender Systems}.
\newblock
\newblock
\showeprint[arxiv]{2004.00646}~[cs.IR]


\bibitem[\protect\citeauthoryear{Lei, Zhang, He, Miao, Wang, Chen, and
  Chua}{Lei et~al\mbox{.}}{2020}]%
        {Lei:2020:IPR}
\bibfield{author}{\bibinfo{person}{Wenqiang Lei}, \bibinfo{person}{Gangyi
  Zhang}, \bibinfo{person}{Xiangnan He}, \bibinfo{person}{Yisong Miao},
  \bibinfo{person}{Xiang Wang}, \bibinfo{person}{Liang Chen}, {and}
  \bibinfo{person}{Tat-Seng Chua}.} \bibinfo{year}{2020}\natexlab{}.
\newblock \showarticletitle{Interactive Path Reasoning on Graph for
  Conversational Recommendation}. In \bibinfo{booktitle}{\emph{Proceedings of
  the 26th ACM SIGKDD International Conference on Knowledge Discovery and Data
  Mining}} \emph{(\bibinfo{series}{KDD '20})}. \bibinfo{pages}{2073–2083}.
\newblock


\bibitem[\protect\citeauthoryear{Lewis, Liu, Goyal, Ghazvininejad, Mohamed,
  Levy, Stoyanov, and Zettlemoyer}{Lewis et~al\mbox{.}}{2020}]%
        {Mike:2019:BART}
\bibfield{author}{\bibinfo{person}{Mike Lewis}, \bibinfo{person}{Yinhan Liu},
  \bibinfo{person}{Naman Goyal}, \bibinfo{person}{Marjan Ghazvininejad},
  \bibinfo{person}{Abdelrahman Mohamed}, \bibinfo{person}{Omer Levy},
  \bibinfo{person}{Veselin Stoyanov}, {and} \bibinfo{person}{Luke
  Zettlemoyer}.} \bibinfo{year}{2020}\natexlab{}.
\newblock \showarticletitle{{BART}: Denoising Sequence-to-Sequence Pre-training
  for Natural Language Generation, Translation, and Comprehension}. In
  \bibinfo{booktitle}{\emph{Proceedings of the 58th Annual Meeting of the
  Association for Computational Linguistics}} \emph{(\bibinfo{series}{ACL
  '20})}. \bibinfo{pages}{7871--7880}.
\newblock


\bibitem[\protect\citeauthoryear{Lin}{Lin}{2004}]%
        {Lin:2004:ROUGE}
\bibfield{author}{\bibinfo{person}{Chin-Yew Lin}.}
  \bibinfo{year}{2004}\natexlab{}.
\newblock \showarticletitle{ROUGE: A Package for Automatic Evaluation of
  Summaries}. In \bibinfo{booktitle}{\emph{Proceedings of the ACL workshop on
  Text Summarization Branches Out}}. \bibinfo{pages}{74--81}.
\newblock


\bibitem[\protect\citeauthoryear{Lipani, Carterette, and Yilmaz}{Lipani
  et~al\mbox{.}}{2021}]%
        {Lipani:2021:TOIS}
\bibfield{author}{\bibinfo{person}{Aldo Lipani}, \bibinfo{person}{Ben
  Carterette}, {and} \bibinfo{person}{Emine Yilmaz}.}
  \bibinfo{year}{2021}\natexlab{}.
\newblock \showarticletitle{How Am I Doing?: Evaluating Conversational Search
  Systems Offline}.
\newblock \bibinfo{journal}{\emph{ACM Trans. Inf. Syst.}} \bibinfo{volume}{39},
  \bibinfo{number}{4}, Article \bibinfo{articleno}{51} (\bibinfo{date}{Aug.}
  \bibinfo{year}{2021}), \bibinfo{numpages}{22}~pages.
\newblock


\bibitem[\protect\citeauthoryear{Lippe, Ren, Haned, Voorn, and de~Rijke}{Lippe
  et~al\mbox{.}}{2021}]%
        {Lippe:2021:DTD}
\bibfield{author}{\bibinfo{person}{Phillip Lippe}, \bibinfo{person}{Pengjie
  Ren}, \bibinfo{person}{Hinda Haned}, \bibinfo{person}{Bart Voorn}, {and}
  \bibinfo{person}{Maarten de Rijke}.} \bibinfo{year}{2021}\natexlab{}.
\newblock \showarticletitle{Diversifying Task-oriented Dialogue Response
  Generation with Prototype Guided Paraphrasing}.
\newblock \bibinfo{journal}{\emph{IEEE/ACM Trans. Audio Speech Lang. Process.}}
  (\bibinfo{year}{2021}).
\newblock
\newblock
\shownote{Submitted.}


\bibitem[\protect\citeauthoryear{Liu, Chen, Lou, Zhou, and Zhang}{Liu
  et~al\mbox{.}}{2020}]%
        {Liu:2020:IUR}
\bibfield{author}{\bibinfo{person}{Qian Liu}, \bibinfo{person}{Bei Chen},
  \bibinfo{person}{Jian-Guang Lou}, \bibinfo{person}{Bin Zhou}, {and}
  \bibinfo{person}{Dongmei Zhang}.} \bibinfo{year}{2020}\natexlab{}.
\newblock \showarticletitle{Incomplete Utterance Rewriting as Semantic
  Segmentation}. In \bibinfo{booktitle}{\emph{Proceedings of the 2020
  Conference on Empirical Methods in Natural Language Processing}}
  \emph{(\bibinfo{series}{EMNLP '20})}. \bibinfo{pages}{2846--2857}.
\newblock


\bibitem[\protect\citeauthoryear{Papangelis, Wang, Molino, and Tur}{Papangelis
  et~al\mbox{.}}{2019}]%
        {Papangelis:2019:CMD}
\bibfield{author}{\bibinfo{person}{Alexandros Papangelis},
  \bibinfo{person}{Yi-Chia Wang}, \bibinfo{person}{Piero Molino}, {and}
  \bibinfo{person}{Gokhan Tur}.} \bibinfo{year}{2019}\natexlab{}.
\newblock \showarticletitle{Collaborative Multi-Agent Dialogue Model Training
  Via Reinforcement Learning}. In \bibinfo{booktitle}{\emph{Proceedings of the
  19th Annual SIGdial Meeting on Discourse and Dialogue}}
  \emph{(\bibinfo{series}{SIGDIAL '19})}. \bibinfo{pages}{92--102}.
\newblock


\bibitem[\protect\citeauthoryear{Papineni, Roukos, Ward, and Zhu}{Papineni
  et~al\mbox{.}}{2002}]%
        {Papineni:2002:BLEU}
\bibfield{author}{\bibinfo{person}{Kishore Papineni}, \bibinfo{person}{Salim
  Roukos}, \bibinfo{person}{Todd Ward}, {and} \bibinfo{person}{Wei-Jing Zhu}.}
  \bibinfo{year}{2002}\natexlab{}.
\newblock \showarticletitle{BLEU: A Method for Automatic Evaluation of Machine
  Translation}. In \bibinfo{booktitle}{\emph{Proceedings of the 40th Annual
  Meeting of the Association for Computational Linguistics}}
  \emph{(\bibinfo{series}{ACL '02})}. \bibinfo{pages}{311–318}.
\newblock


\bibitem[\protect\citeauthoryear{Peng, Li, Gao, Liu, and Wong}{Peng
  et~al\mbox{.}}{2018}]%
        {Peng:2018:DIP}
\bibfield{author}{\bibinfo{person}{Baolin Peng}, \bibinfo{person}{Xiujun Li},
  \bibinfo{person}{Jianfeng Gao}, \bibinfo{person}{Jingjing Liu}, {and}
  \bibinfo{person}{Kam-Fai Wong}.} \bibinfo{year}{2018}\natexlab{}.
\newblock \showarticletitle{{D}eep {D}yna-{Q}: Integrating Planning for
  Task-Completion Dialogue Policy Learning}. In
  \bibinfo{booktitle}{\emph{Proceedings of the 56th Annual Meeting of the
  Association for Computational Linguistics}} \emph{(\bibinfo{series}{ACL
  '18})}. \bibinfo{pages}{2182--2192}.
\newblock


\bibitem[\protect\citeauthoryear{Qu, Yang, Croft, Trippas, Zhang, and Qiu}{Qu
  et~al\mbox{.}}{2018}]%
        {Qu:2018:ACU}
\bibfield{author}{\bibinfo{person}{Chen Qu}, \bibinfo{person}{Liu Yang},
  \bibinfo{person}{W.~Bruce Croft}, \bibinfo{person}{Johanne~R. Trippas},
  \bibinfo{person}{Yongfeng Zhang}, {and} \bibinfo{person}{Minghui Qiu}.}
  \bibinfo{year}{2018}\natexlab{}.
\newblock \showarticletitle{Analyzing and Characterizing User Intent in
  Information-Seeking Conversations}. In \bibinfo{booktitle}{\emph{Proceedings
  of the 41st International ACM SIGIR Conference on Research and Development in
  Information Retrieval}} \emph{(\bibinfo{series}{SIGIR '18})}.
  \bibinfo{pages}{989--992}.
\newblock


\bibitem[\protect\citeauthoryear{Radford, Wu, Child, Luan, Amodei, and
  Sutskever}{Radford et~al\mbox{.}}{2019}]%
        {Radford:2019:GPT2}
\bibfield{author}{\bibinfo{person}{Alec Radford}, \bibinfo{person}{Jeff Wu},
  \bibinfo{person}{Rewon Child}, \bibinfo{person}{David Luan},
  \bibinfo{person}{Dario Amodei}, {and} \bibinfo{person}{Ilya Sutskever}.}
  \bibinfo{year}{2019}\natexlab{}.
\newblock \showarticletitle{Language Models are Unsupervised Multitask
  Learners}.
\newblock  (\bibinfo{year}{2019}).
\newblock


\bibitem[\protect\citeauthoryear{Radlinski, Balog, Byrne, and
  Krishnamoorthi}{Radlinski et~al\mbox{.}}{2019}]%
        {Radlinski:2019:CCP}
\bibfield{author}{\bibinfo{person}{Filip Radlinski}, \bibinfo{person}{Krisztian
  Balog}, \bibinfo{person}{Bill Byrne}, {and} \bibinfo{person}{Karthik
  Krishnamoorthi}.} \bibinfo{year}{2019}\natexlab{}.
\newblock \showarticletitle{Coached Conversational Preference Elicitation: {A}
  Case Study in Understanding Movie Preferences}. In
  \bibinfo{booktitle}{\emph{Proceedings of the 20th Annual SIGdial Meeting on
  Discourse and Dialogue}} \emph{(\bibinfo{series}{SIGDIAL '19})}.
  \bibinfo{pages}{353--360}.
\newblock


\bibitem[\protect\citeauthoryear{Radlinski and Craswell}{Radlinski and
  Craswell}{2017}]%
        {Radlinski:2017:TFC}
\bibfield{author}{\bibinfo{person}{Filip Radlinski} {and} \bibinfo{person}{Nick
  Craswell}.} \bibinfo{year}{2017}\natexlab{}.
\newblock \showarticletitle{A Theoretical Framework for Conversational Search}.
  In \bibinfo{booktitle}{\emph{Proceedings of the 2017 Conference on Conference
  Human Information Interaction and Retrieval}} \emph{(\bibinfo{series}{CHIIR
  '17})}. \bibinfo{pages}{117--126}.
\newblock


\bibitem[\protect\citeauthoryear{Raffel, Shazeer, Roberts, Lee, Narang, Matena,
  Zhou, Li, and Liu}{Raffel et~al\mbox{.}}{2019}]%
        {Colin:2019:T5}
\bibfield{author}{\bibinfo{person}{Colin Raffel}, \bibinfo{person}{Noam
  Shazeer}, \bibinfo{person}{Adam Roberts}, \bibinfo{person}{Katherine Lee},
  \bibinfo{person}{Sharan Narang}, \bibinfo{person}{Michael Matena},
  \bibinfo{person}{Yanqi Zhou}, \bibinfo{person}{Wei Li}, {and}
  \bibinfo{person}{Peter~J. Liu}.} \bibinfo{year}{2019}\natexlab{}.
\newblock \bibinfo{title}{Exploring the Limits of Transfer Learning with a
  Unified Text-to-Text Transformer}.
\newblock
\newblock
\showeprint[arxiv]{1910.10683}~[cs.LG]


\bibitem[\protect\citeauthoryear{Rha, Shi, and Belkin}{Rha
  et~al\mbox{.}}{2017}]%
        {Rha:2017:ERQ}
\bibfield{author}{\bibinfo{person}{Eun Rha}, \bibinfo{person}{Wei Shi}, {and}
  \bibinfo{person}{Nicholas Belkin}.} \bibinfo{year}{2017}\natexlab{}.
\newblock \showarticletitle{An Exploration of Reasons for Query
  Reformulations}.
\newblock \bibinfo{journal}{\emph{J. Assoc. Inf. Sci. Technol}}
  \bibinfo{volume}{54} (\bibinfo{year}{2017}), \bibinfo{pages}{337--346}.
\newblock


\bibitem[\protect\citeauthoryear{Rosset, Xiong, Song, Campos, Craswell, Tiwary,
  and Bennett}{Rosset et~al\mbox{.}}{2020}]%
        {Rosset:2020:LCS}
\bibfield{author}{\bibinfo{person}{Corbin Rosset}, \bibinfo{person}{Chenyan
  Xiong}, \bibinfo{person}{Xia Song}, \bibinfo{person}{Daniel Campos},
  \bibinfo{person}{Nick Craswell}, \bibinfo{person}{Saurabh Tiwary}, {and}
  \bibinfo{person}{Paul Bennett}.} \bibinfo{year}{2020}\natexlab{}.
\newblock \showarticletitle{Leading Conversational Search by Suggesting Useful
  Questions}. In \bibinfo{booktitle}{\emph{Proceedings of The Web Conference
  2020}} \emph{(\bibinfo{series}{WWW '20})}. \bibinfo{pages}{1160--1170}.
\newblock


\bibitem[\protect\citeauthoryear{Salle, Malmasi, Rokhlenko, and
  Agichtein}{Salle et~al\mbox{.}}{2021}]%
        {Salle:2021:SEC}
\bibfield{author}{\bibinfo{person}{Alexandre Salle}, \bibinfo{person}{Shervin
  Malmasi}, \bibinfo{person}{Oleg Rokhlenko}, {and} \bibinfo{person}{Eugene
  Agichtein}.} \bibinfo{year}{2021}\natexlab{}.
\newblock \showarticletitle{Studying the Effectiveness of Conversational Search
  Refinement Through User Simulation}. In \bibinfo{booktitle}{\emph{Proceedings
  of the 43rd European Conference on Information Retrieval}}
  \emph{(\bibinfo{series}{ECIR '21})}. \bibinfo{pages}{587--602}.
\newblock


\bibitem[\protect\citeauthoryear{Schatzmann, Thomson, Weilhammer, Ye, and
  Young}{Schatzmann et~al\mbox{.}}{2007}]%
        {Schatzmann:2007:AUS}
\bibfield{author}{\bibinfo{person}{Jost Schatzmann}, \bibinfo{person}{Blaise
  Thomson}, \bibinfo{person}{Karl Weilhammer}, \bibinfo{person}{Hui Ye}, {and}
  \bibinfo{person}{Steve Young}.} \bibinfo{year}{2007}\natexlab{}.
\newblock \showarticletitle{Agenda-based User Simulation for Bootstrapping a
  POMDP Dialogue System}. In \bibinfo{booktitle}{\emph{Proceedings of the Human
  Language Technology Conference of the North American Chapter of the
  {A}ssociation for Computational Linguistics}}
  \emph{(\bibinfo{series}{HLT-NAACL '07})}. \bibinfo{pages}{149--152}.
\newblock


\bibitem[\protect\citeauthoryear{Sekulic, Aliannejadi, and Crestani}{Sekulic
  et~al\mbox{.}}{2022}]%
        {Sekulic:2022:EMC}
\bibfield{author}{\bibinfo{person}{Ivan Sekulic}, \bibinfo{person}{Mohammad
  Aliannejadi}, {and} \bibinfo{person}{Fabio Crestani}.}
  \bibinfo{year}{2022}\natexlab{}.
\newblock \showarticletitle{Evaluating Mixed-initiative Conversational Search
  Systems via User Simulation}. In \bibinfo{booktitle}{\emph{The 15th
  International Conference on Web Search and Data Mining}}
  \emph{(\bibinfo{series}{WSDM '22})}. \bibinfo{pages}{888–896}.
\newblock


\bibitem[\protect\citeauthoryear{Su, Shen, Zhang, Sun, Hu, Niu, and Zhou}{Su
  et~al\mbox{.}}{2019}]%
        {Su:2019:IMD}
\bibfield{author}{\bibinfo{person}{Hui Su}, \bibinfo{person}{Xiaoyu Shen},
  \bibinfo{person}{Rongzhi Zhang}, \bibinfo{person}{Fei Sun},
  \bibinfo{person}{Pengwei Hu}, \bibinfo{person}{Cheng Niu}, {and}
  \bibinfo{person}{Jie Zhou}.} \bibinfo{year}{2019}\natexlab{}.
\newblock \showarticletitle{Improving Multi-turn Dialogue Modelling with
  Utterance {R}e{W}riter}. In \bibinfo{booktitle}{\emph{Proceedings of the 57th
  Annual Meeting of the Association for Computational Linguistics}}
  \emph{(\bibinfo{series}{ACL '19})}. \bibinfo{pages}{22--31}.
\newblock


\bibitem[\protect\citeauthoryear{Sun, Zhang, Balog, Ren, Ren, Chen, and
  de~Rijke}{Sun et~al\mbox{.}}{2021}]%
        {Sun:2021:SUS}
\bibfield{author}{\bibinfo{person}{Weiwei Sun}, \bibinfo{person}{Shuo Zhang},
  \bibinfo{person}{Krisztian Balog}, \bibinfo{person}{Zhaochun Ren},
  \bibinfo{person}{Pengjie Ren}, \bibinfo{person}{Zhumin Chen}, {and}
  \bibinfo{person}{Maarten de Rijke}.} \bibinfo{year}{2021}\natexlab{}.
\newblock \showarticletitle{Simulating User Satisfaction for the Evaluation of
  Task-Oriented Dialogue Systems}. In \bibinfo{booktitle}{\emph{Proceedings of
  the 44th International ACM SIGIR Conference on Research and Development in
  Information Retrieval}} \emph{(\bibinfo{series}{SIGIR '21})}.
  \bibinfo{pages}{2499--2506}.
\newblock


\bibitem[\protect\citeauthoryear{Sun and Zhang}{Sun and Zhang}{2018}]%
        {Sun:2018:CRS}
\bibfield{author}{\bibinfo{person}{Yueming Sun} {and} \bibinfo{person}{Yi
  Zhang}.} \bibinfo{year}{2018}\natexlab{}.
\newblock \showarticletitle{Conversational Recommender System}. In
  \bibinfo{booktitle}{\emph{The 41st International ACM SIGIR Conference on
  Research and Development in Information Retrieval}}
  \emph{(\bibinfo{series}{SIGIR '18})}. \bibinfo{pages}{235--244}.
\newblock


\bibitem[\protect\citeauthoryear{Teodorescu}{Teodorescu}{2009}]%
        {Iuliana:2009:MLE}
\bibfield{author}{\bibinfo{person}{Iuliana Teodorescu}.}
  \bibinfo{year}{2009}\natexlab{}.
\newblock \bibinfo{title}{Maximum Likelihood Estimation for Markov Chains}.
\newblock
\newblock
\showeprint[arxiv]{0905.4131}~[stat.CO]


\bibitem[\protect\citeauthoryear{Trippas, Spina, Cavedon, Joho, and
  Sanderson}{Trippas et~al\mbox{.}}{2018}]%
        {Trippas:2018:IDS}
\bibfield{author}{\bibinfo{person}{Johanne~R. Trippas},
  \bibinfo{person}{Damiano Spina}, \bibinfo{person}{Lawrence Cavedon},
  \bibinfo{person}{Hideo Joho}, {and} \bibinfo{person}{Mark Sanderson}.}
  \bibinfo{year}{2018}\natexlab{}.
\newblock \showarticletitle{Informing the Design of Spoken Conversational
  Search: Perspective Paper}. In \bibinfo{booktitle}{\emph{Proceedings of the
  2018 Conference on Human Information Interaction and Retrieval}}
  \emph{(\bibinfo{series}{CHIIR '18})}. \bibinfo{pages}{32--41}.
\newblock


\bibitem[\protect\citeauthoryear{Trippas, Spina, Scholer, Awadallah, Bailey,
  Bennett, White, Liono, Ren, Salim, and Sanderson}{Trippas
  et~al\mbox{.}}{2019}]%
        {Trippas:2019:LAW}
\bibfield{author}{\bibinfo{person}{Johanne~R. Trippas},
  \bibinfo{person}{Damiano Spina}, \bibinfo{person}{Falk Scholer},
  \bibinfo{person}{Ahmed~Hassan Awadallah}, \bibinfo{person}{Peter Bailey},
  \bibinfo{person}{Paul~N. Bennett}, \bibinfo{person}{Ryen~W. White},
  \bibinfo{person}{Jonathan Liono}, \bibinfo{person}{Yongli Ren},
  \bibinfo{person}{Flora~D. Salim}, {and} \bibinfo{person}{Mark Sanderson}.}
  \bibinfo{year}{2019}\natexlab{}.
\newblock \showarticletitle{Learning About Work Tasks to Inform Intelligent
  Assistant Design}. In \bibinfo{booktitle}{\emph{Proceedings of the 2019
  Conference on Human Information Interaction and Retrieval}}
  \emph{(\bibinfo{series}{CHIIR '19})}. \bibinfo{pages}{5--14}.
\newblock


\bibitem[\protect\citeauthoryear{Vakulenko, Longpre, Tu, and Anantha}{Vakulenko
  et~al\mbox{.}}{2021}]%
        {Vakulenko:2020:QRF}
\bibfield{author}{\bibinfo{person}{Svitlana Vakulenko}, \bibinfo{person}{Shayne
  Longpre}, \bibinfo{person}{Zhucheng Tu}, {and} \bibinfo{person}{Raviteja
  Anantha}.} \bibinfo{year}{2021}\natexlab{}.
\newblock \showarticletitle{Question Rewriting for Conversational Question
  Answering}. In \bibinfo{booktitle}{\emph{Proceedings of the 14th ACM
  International Conference on Web Search and Data Mining}}
  \emph{(\bibinfo{series}{WSDM '21})}. \bibinfo{pages}{355--363}.
\newblock


\bibitem[\protect\citeauthoryear{Vtyurina, Clarke, Law, Trippas, and
  Bota}{Vtyurina et~al\mbox{.}}{2020}]%
        {Vtyurina:2020:AMA}
\bibfield{author}{\bibinfo{person}{Alexandra Vtyurina},
  \bibinfo{person}{Charles L.~A. Clarke}, \bibinfo{person}{Edith Law},
  \bibinfo{person}{Johanne~R. Trippas}, {and} \bibinfo{person}{Horatiu Bota}.}
  \bibinfo{year}{2020}\natexlab{}.
\newblock \showarticletitle{A Mixed-Method Analysis of Text and Audio Search
  Interfaces with Varying Task Complexity}. In
  \bibinfo{booktitle}{\emph{Proceedings of the 2020 ACM SIGIR on International
  Conference on Theory of Information Retrieval}} \emph{(\bibinfo{series}{ICTIR
  '20})}. \bibinfo{pages}{61--68}.
\newblock


\bibitem[\protect\citeauthoryear{Vtyurina and Fourney}{Vtyurina and
  Fourney}{2018}]%
        {Vtyurina:2018:ERC}
\bibfield{author}{\bibinfo{person}{Alexandra Vtyurina} {and}
  \bibinfo{person}{Adam Fourney}.} \bibinfo{year}{2018}\natexlab{}.
\newblock \showarticletitle{Exploring the Role of Conversational Cues in Guided
  Task Support with Virtual Assistants}. In
  \bibinfo{booktitle}{\emph{Proceedings of the 2018 CHI Conference on Human
  Factors in Computing Systems}} \emph{(\bibinfo{series}{CHI '18})}.
  \bibinfo{pages}{1--7}.
\newblock


\bibitem[\protect\citeauthoryear{Vtyurina, Savenkov, Agichtein, and
  Clarke}{Vtyurina et~al\mbox{.}}{2017}]%
        {Vtyurina:2017:ECS}
\bibfield{author}{\bibinfo{person}{Alexandra Vtyurina}, \bibinfo{person}{Denis
  Savenkov}, \bibinfo{person}{Eugene Agichtein}, {and} \bibinfo{person}{Charles
  L.~A. Clarke}.} \bibinfo{year}{2017}\natexlab{}.
\newblock \showarticletitle{Exploring Conversational Search With Humans,
  Assistants, and Wizards}. In \bibinfo{booktitle}{\emph{Proceedings of the
  2017 CHI Conference Extended Abstracts on Human Factors in Computing
  Systems}} \emph{(\bibinfo{series}{CHI EA '17})}. \bibinfo{pages}{2187--2193}.
\newblock


\bibitem[\protect\citeauthoryear{Wan}{Wan}{2018}]%
        {Wan:2018:ATS}
\bibfield{author}{\bibinfo{person}{Xiaojun Wan}.}
  \bibinfo{year}{2018}\natexlab{}.
\newblock \showarticletitle{Automatic Text Simplification by Horacio Saggion}.
\newblock \bibinfo{journal}{\emph{J. Comput. Linguist}} \bibinfo{volume}{44},
  \bibinfo{number}{4} (\bibinfo{year}{2018}), \bibinfo{pages}{659--661}.
\newblock


\bibitem[\protect\citeauthoryear{Warstadt, Singh, and Bowman}{Warstadt
  et~al\mbox{.}}{2018}]%
        {Alex:2018:NNAJ}
\bibfield{author}{\bibinfo{person}{Alex Warstadt}, \bibinfo{person}{Amanpreet
  Singh}, {and} \bibinfo{person}{Samuel~R. Bowman}.}
  \bibinfo{year}{2018}\natexlab{}.
\newblock \bibinfo{title}{Neural Network Acceptability Judgments}.
\newblock
\newblock
\showeprint[arxiv]{1805.12471}~[cs.CL]


\bibitem[\protect\citeauthoryear{Wolf, Debut, Sanh, Chaumond, Delangue, Moi,
  Cistac, Rault, Louf, Funtowicz, Davison, Shleifer, von Platen, Ma, Jernite,
  Plu, Xu, Le~Scao, Gugger, Drame, Lhoest, and Rush}{Wolf
  et~al\mbox{.}}{2020}]%
        {Thomas:2019:HT}
\bibfield{author}{\bibinfo{person}{Thomas Wolf}, \bibinfo{person}{Lysandre
  Debut}, \bibinfo{person}{Victor Sanh}, \bibinfo{person}{Julien Chaumond},
  \bibinfo{person}{Clement Delangue}, \bibinfo{person}{Anthony Moi},
  \bibinfo{person}{Pierric Cistac}, \bibinfo{person}{Tim Rault},
  \bibinfo{person}{Remi Louf}, \bibinfo{person}{Morgan Funtowicz},
  \bibinfo{person}{Joe Davison}, \bibinfo{person}{Sam Shleifer},
  \bibinfo{person}{Patrick von Platen}, \bibinfo{person}{Clara Ma},
  \bibinfo{person}{Yacine Jernite}, \bibinfo{person}{Julien Plu},
  \bibinfo{person}{Canwen Xu}, \bibinfo{person}{Teven Le~Scao},
  \bibinfo{person}{Sylvain Gugger}, \bibinfo{person}{Mariama Drame},
  \bibinfo{person}{Quentin Lhoest}, {and} \bibinfo{person}{Alexander Rush}.}
  \bibinfo{year}{2020}\natexlab{}.
\newblock \showarticletitle{Transformers: State-of-the-Art Natural Language
  Processing}. In \bibinfo{booktitle}{\emph{Proceedings of the 2020 Conference
  on Empirical Methods in Natural Language Processing: System Demonstrations}}
  \emph{(\bibinfo{series}{EMNLP '20})}. \bibinfo{pages}{38--45}.
\newblock


\bibitem[\protect\citeauthoryear{Zamani, Trippas, Dalton, and Radlinski}{Zamani
  et~al\mbox{.}}{2022}]%
        {Zamani:2022:FnTIR}
\bibfield{author}{\bibinfo{person}{Hamed Zamani}, \bibinfo{person}{Johanne~R.
  Trippas}, \bibinfo{person}{Jeff Dalton}, {and} \bibinfo{person}{Filip
  Radlinski}.} \bibinfo{year}{2022}\natexlab{}.
\newblock \bibinfo{title}{Conversational Information Seeking}.
\newblock
\newblock
\showeprint[arxiv]{2201.08808}~[cs.IR]


\bibitem[\protect\citeauthoryear{Zhang and Balog}{Zhang and Balog}{2020}]%
        {Zhang:2020:ECR}
\bibfield{author}{\bibinfo{person}{Shuo Zhang} {and} \bibinfo{person}{Krisztian
  Balog}.} \bibinfo{year}{2020}\natexlab{}.
\newblock \showarticletitle{Evaluating Conversational Recommender Systems via
  User Simulation}. In \bibinfo{booktitle}{\emph{Proceedings of the 26th ACM
  SIGKDD International Conference on Knowledge Discovery and Data Mining}}
  \emph{(\bibinfo{series}{KDD '20})}. \bibinfo{pages}{1512--1520}.
\newblock


\end{thebibliography}


\end{document}